\documentclass[10pt,journal,compsoc]{IEEEtran}

\ifCLASSINFOpdf
\else
\fi

\ifCLASSOPTIONcompsoc
  \usepackage[nocompress]{cite}
\else
  \usepackage{cite}
\fi

\usepackage{amssymb}
\usepackage{amsmath}
\usepackage{algorithm}
\usepackage[noend]{algorithmic}
\usepackage{graphicx}
\usepackage{subfigure}
\usepackage{xcolor}
\usepackage{epstopdf}
\usepackage{epsfig}
\usepackage{multirow}

\newtheorem{myexp}{Example}

\begin{document}

\title{Regular Expression Matching on billion-nodes Graphs}


%
%
%

\author{Hongzhi~Wang,~\IEEEmembership{Member,~IEEE,}
        Jiabao~Han,
        and,
        Bin Shao,~Jianzhong~Li,~\IEEEmembership{Member,~IEEE}
\IEEEcompsocitemizethanks{\IEEEcompsocthanksitem Hongzhi Wang, Jiabao Han and Jianzhong Li are with the Department
of Computer Science and Technology, Harbin Institute of Technology, China.\protect\\
E-mail: wangzh@hit.edu.cn
\IEEEcompsocthanksitem Bin Shao is with Microsoft Research Asia.}
}


\date{}

\IEEEtitleabstractindextext{
\begin{abstract}
In many applications, it is necessary to retrieve pairs of vertices
with the path between them satisfying certain constraints, since
regular expression is a powerful tool to describe patterns of a
sequence. To meet such requirements, in this paper, we define regular
expression (RE) query on graphs to use regular expression to represent
the constraints between vertices. To process RE queries on large
graphs such as social networks, we propose the RE query processing
method with the index size sublinear to the graph size. Considering
that large graphs may be randomly distributed in multiple machines, the parallel RE processing algorithms are presented
without the assumption of graph distribution. To achieve high
efficiency for complex RE query processing, we develop cost-based
query optimization strategies with only a small size statistical
information which is suitable for querying large graphs.
Comprehensive experimental results show that this approach works scale well
for large graphs.
\end{abstract}
}

\maketitle

\IEEEdisplaynontitleabstractindextext

\IEEEpeerreviewmaketitle

\IEEEraisesectionheading{\section{Introduction}}
\label{sec:intro}

Graph data have been widely applied in many area such as knowledge
management~\cite{DBLP:journals/pvldb/ZengYWSW13}, social
network~\cite{facebook}, bioinformatics~\cite{debruijn}, and
compilers~\cite{DBLP:conf/fase/ZhangZL11}.

In the area, an important application is to retrieve pairs of vertices
with specific path between them. For instance, in knowledge management, the information of whether two persons with common characteristics come from the same lineage may be required to extract. It is hard to express by a subgraph or SPARQL query.

As a powerful tool of string pattern description, regular expression
can also be used to describe the relationship of vertices in a
graph. RE allows querying of arbitrary length paths by using regular
expression patterns, it is useful for expressing complex navigation in
a graph, in particular, union and transitive closure are crucial when
one does not have a complete knowledge of the structure of the
knowledge base. In our example, suppose a person $M$ and a person $N$ share a common characteristic. In other words, they share a common attribute values in the knowledge database. If we want to know whether $M$ and $N$ share to the same lineage. This query can be expressed by regular expression easily.

Even some existing work~\cite{DBLP:conf/icde/FanLMTW11,DBLP:conf/vldb/LiM01,DBLP:conf/popl/HosoyaP01,DBLP:journals/toit/HosoyaP03,DBLP:conf/edbt/YakovetsGG15,DBLP:conf/edbt/FletcherPP16,DBLP:conf/sigmod/YakovetsGG16} start to study regular expression on graphs, their methods have two shortcomings.

One one hand, existing approaches
only consider a share of operations in regular expressions. \cite{DBLP:conf/icde/FanLMTW11} only studied
the regular expression matching on the graph without ``or''($|$) and
recursion closure.
\cite{DBLP:conf/vldb/LiM01,DBLP:conf/popl/HosoyaP01,DBLP:journals/toit/HosoyaP03}
considers only the regular expression matching on tree structure
data. Thus, the range of application for these methods is limited.

On
the other hand, the scalability is not considered sufficiently.
Currently, a graph may scale to very large, even to billion-nodes.
For example, FreeBase, the online query processing and services on frequently updating
graphs with large size requires lightweight indices. However, \cite{DBLP:conf/icde/FanLMTW11} requires the structural index
with size at least the number of edges of graph. Even though it
works efficiently on graphs with small size, it is not suitable for
large graphs. \cite{DBLP:conf/edbt/YakovetsGG15,DBLP:conf/edbt/FletcherPP16,DBLP:conf/sigmod/YakovetsGG16} do not consider the parallelization of regular expression matching, and they could hardly handle very large graphs that could not reside in a single machine.

To process flexible regular expression on billion-nodes graphs with only
a lightweight index, in this paper, a search-based method is
proposed. In this method, only an index that retrieves the vertices
by their labels is used, which is in size linear to the number of nodes and easy to update. Without structural index, our method is
traversal-based.

To avoid large intermediate results, in our system,
a lightweight representation of intermediate results is designed.
Based on such representation, we propose the basic operators for
regular expression processing. For the scalability issues,
besides implementation in a single computer, we propose parallel
implementation of the operators in a cluster. To find an efficient
execution order of the basic operators, a cost-based query
optimization strategy for regular expression is proposed with a cost model
which requires only small statistics information of graphs. The contributions
of this paper are summarized as follows.

\begin{list}{\labelitemi}{\leftmargin=1em}\itemsep 0pt \parskip 0pt
	\item The regular expression query processing on a large graph with full features
	is studied in this paper. To the best of our knowledge, this is the first
	paper that studies parallel processing of the full features of regular expression on large graph
	data.
	
	\item With the consideration of the requirement of large graph, we
	propose a traversal-based strategy for regular expression processing
	on a large graph. For the convenience of query optimization, we
	divide query processing into operators. For each operator,
	efficient implementation algorithms on both single computer and
	clusters are developed.
	
	\item For query optimization of regular expression queries on graph,
	a cost model is proposed. With the consideration of the feature of
	large graphs, an estimation strategy requiring only simple statistic
	information is proposed.
	
	\item A dynamic programming algorithm is designed to obtain
	efficient execution order of operators for regular expression
	processing based on the estimated cost.
	
	\item Extensive experimental results demonstrate that our approaches could accomplish regular expression matching on billion-nodes graphs within 2 seconds and are suitable for various graphs and queries.
\end{list}

Section~\ref{sec:def} describes the query language. In Section~\ref{sec:framework}, the framework of
regular expression processing is proposed. The logical operator generation is described in Section~\ref{sec:LoOp}. The implementation
algorithms of the operators are presented in
Section~\ref{sec:operator}. The cost model and query optimization
strategy are described in Section~\ref{sec:opt}. Section~\ref{sec:extension} discusses extension issues of the proposed approaches. Experiments and analysis are described in Section~\ref{sec:exper}.
Section~\ref{sec:related} summaries related work and
Section~\ref{sec:con} concludes the whole paper.

\section{Query Language}
\label{sec:def}

In many scenarios, the relationship between two vertices in a graph may be complex. In a knowledge base, the relationship between vertices could be naturally represented as the label sequence on the path between them. Thus, to retrieve vertex pairs satisfying requirements in a graph, the query language should describe the label sequence required to be between two vertices. However, with the increase of path length, the relationship between two vertices becomes too complex to be expressed with compact queries with simple semantics.

For instance, $N$ is the offspring of $M$. Obviously, they share the same lineage. In the specific case, we can hardly obtain the length of the path between $M$ and $N$. Thus, it is difficult to express this relationship.

Regular expression is a form of language for sequence in common use. Similar as the regular expression for strings, a regular expression can
describe labels attached to the vertices in a path
between two vertices in a labeled graph. Hence, a regular expression
can be used to describe the requirement of the paths between two
vertices in a graph. In the above example, each person has a property `status', it has a label $p$ if he or she has a child and everyone has a label $c$. If $M$ and $N$ come from the same family, the path between $M$ and $N$ must satisfy a criteria: $p-c-p-c-p-c\ldots\ldots$, where we use `-' to denote concatenation to avoid confusion. If we use regular expression to express this criteria, it can be expressed like this: $p(c-p)^+c$. Note that reachability queries cannot express such query properly since reachability queries neglect the label in the path.

Formally, the regular
expression is defined as following. For the flexibility, the
wildcard representing any label is added to the definition.

Suppose the tag set is $\Sigma$. The wildcard is denoted by ``\#''.
The regular expression on $\Sigma$, denoted by $RE(\Sigma)$, is
defined as follows.

\begin{list}{\labelitemi}{\leftmargin=1em}\itemsep 0pt \parskip 0pt
	\item $\Sigma\cup \{\#\}\subseteq RE(\Sigma)$
	
	\item $\forall r_1, r_2\in RE(\Sigma)$, $r_1r_2\in RE(\Sigma)$
	
	\item $\forall r_1, r_2\in RE(\Sigma)$, $r_1|r_2\in RE(\Sigma)$
	
	\item $\forall r\in RE(\Sigma)$, $r^+\in  RE(\Sigma)$
	
	\item No other expressions are in $RE(\Sigma)$
	
\end{list}

A data graph is defined as a graph $G$=($V_G$, $E_G$, $T_V$), where
$V_G$ is the vertex set, $E_G$ is the edge set, and $T_V:V_G\rightarrow \Sigma\cup `\#'$ assigns a label to each vertex in $V_G$.

If the string $s_P$ constructed by connecting the tags in the
vertices of a path $P$ from the start node to the end node matches a
regular expression $E$, it is said that $P$ matches $E$.

The result of a regular expression $E$ on a data graph $G$ is a set
of pairs $R_{E,G}$=\{($(v_1, v_k$)$|$ a path
($v_1$$\rightarrow$$v_2$$\rightarrow$$\cdots$$\rightarrow$$v_k$)
exists in $G$ matches $E$\}.

\begin{figure}
	\centering
	\includegraphics[width=0.3\linewidth]{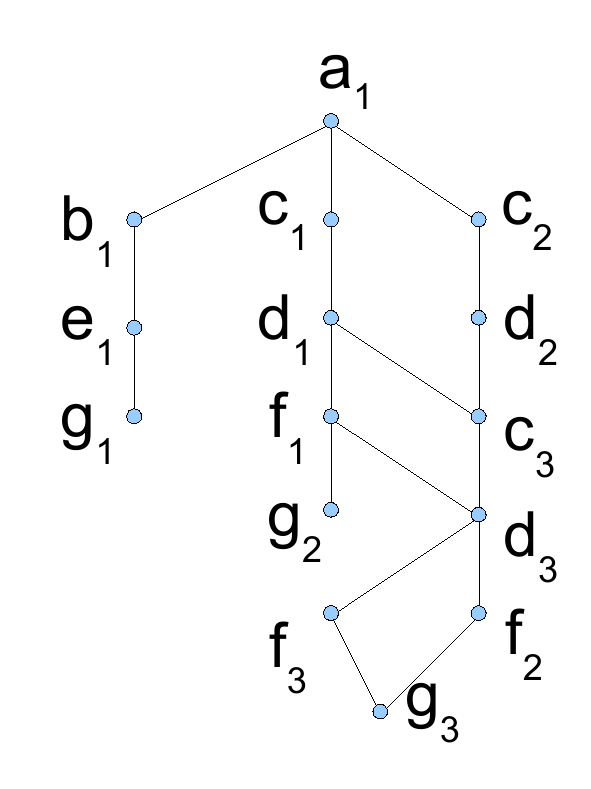}
	\caption{The Data Graph}  \label{fig:data_graph}
\end{figure}

For example, consider the data graph shown in
Figure~\ref{fig:data_graph}. The results of the RE query
$Q_1$=$a(be|(cd)^+f)g$ are \{($a_1$, $g_1$), ($a_1$, $g_2$), ($a_1$,
$g_3$)\}. For $a_1$ and $g_1$, the path between them
$a_1b_1e_1g_1$ match the path $abeg$ of the RE. The path
$a_1c_2d_2c_3d_3f_2g_3$ match the path $a(cd)^+fg$ in the RE.

%
%

\section{The Framework of RE Query Processing}
\label{sec:framework}
In this section, we present the framework of RE query processing. Since our goal is to process RE query on large graphs, only lightweight indices with size linear or sublinear to the graph size are permitted. As discussed in \cite{DBLP:journals/pvldb/SunWWSL12}, with only lightweight index, only simple operators are supported for structural query processing, e.g. traversal and join. RE query processing is to combine such simple operators. Thus, we attempt to process a RE query $r$ by separating $r$ into segments properly, each of which is processed by traversal separately, and joining the results of these segments.

With the consideration of the complexity of RE query, it is difficult to perform the query splitting directly. Thus, we handle the query in two tiers. The first one converts a query to the logical plan, which is near to the form of the query. The second one is to generate the physical plan from the logical plan. Then we discuss these two tiers, respectively.

\subsection{Logical Operators}

Since a RE includes three basic operations, concentration, alternative operator $|$ and
closure operation:$^+$, it is natural to match regular expression by turning these
basic operations to the matching in a graph. Thus, we define logical
operators to support the matching according to the operations in RE.

In RE, the operations \textsf{concentration} and \textsf{closure} are used to describe
the connection relationships between nodes in the graph. \textsf{Concentration} means the directly connection between two nodes in a path. \textsf{Closure} means the emergence of node sequences with same label sequence in a path consecutively and repeatedly. The logical
operators for RE are defined according to these two operations. To support the operation of $'|'$, each variable in the
operator is permitted to match multiple labels in the RE.

The input of the operators may be the nodes matching some labels in the RE or the intermediate results matching a sub expression.
With the consideration of input source, corresponding operation in RE and search direction, we summarize 6 logical operators as shown in
Table~\ref{tab:logical_operator}, where each lowercase letter is a
variable referring to a set of labels in RE. In this table, we use
`-' to represent `$\rightarrow$' or `$\leftarrow$', which identify
the direction of the execution.

\begin{table*}
	\caption{The Table of Logical Operators}
	\label{tab:logical_operator}
	\centering
	\begin{tabular}{r|l|p{12cm}}
		\hline
		Operator & Logical Operator  & Semantics\\
		\hline
		concentration & $a-b$& directly concentration between an $a$ node and a $b$ node\\
		& $a-E(b)$ & concentration between a $a$ node and the $b$ node in the result of
		a RE $E$\\
		& $E(b)-a$ & concentration between the $b$ node in result of a RE and a $a$
		node\\
		& $E_1(a)-E_2(b)$ & concentration of the results of
		$E_1$ and $E_2$ on the $a$ and $b$ nodes\\
		\hline
		Closure & $E(E_1(a)-E_2(b))^+$ &  the closure of a RE $E$ whose head is an $a$ node as the tail of $E_1$, tail is a $b$ node as the head of $E_2$, where $E_1$ and $E_2$ have been processed\\
		& $a^+$ & the closure of a single label $a$\\
		\hline
	\end{tabular}
\end{table*}

Intuitively, the processing of an RE query can be converted to a series of
logical operators. The logical operators in a regular expression may have
multiple possible execution orders. How to generate
an efficient execution order is the task of query optimization, wich will be discussed in
Section~\ref{sec:opt}.

Example~\ref{exp:framework} demonstrates the semantics of the
logical operators and the framework of regular expression query
processing.

\begin{myexp}
	\label{exp:framework}
	Consider the query $a(be|(cd)^+f)g$ that is processed in from left to right. The first operator is
	$a\rightarrow (b|c)(\{b,c\})$. The results of this operator are the set of path fragments, each of which matches $a-b$ or $a-c$. Here $b|c$ is a regular expression. $(\{b,c\})$ means that $a$ could be connected to a $b$ or $c$ node.
	
	The following operator is
	$a(b|c)(\{b\})
	\rightarrow e$. It means that the $b$ node in a result of $a(b|c)$ connects to an $e$ node.
	In Figure~\ref{fig:data_graph},
	the results of the subquery $a(b|c)$ are \{($a_1$, $b_1$),($a_1$,
	$c_1$),($a_1$, $c_2$)\}. The partial results of $a(b|c)(\{b\})
	\rightarrow e$ are \{($a_1$, $e_1$),($a_1$,
	$c_1$),($a_1$, $c_2$)\}, corresponding to the regular expression
	$E_1=a(be|c)$.
	
	The following operator is closure $(c\rightarrow
	d)^+(E_1(c),d)$. The intermediate results in this step are $\{(a_1$,
	$e_1$), ($a_1$, $d_1$), ($a_1$, $d_2$), ($a_1$, $d_3)\}$ as the results of
	partial query $E_2=a(be|(cd)^+)$. By processing the operator
	$E_2(d)\rightarrow f$, the intermediate results \{($a_1$, $e_1$), ($a_1$,
	$f_1$), ($a_1$, $f_2$), ($a_1$, $f_3)\}$ are obtained for the partial query
	$E_3=a(be|(cd)^+f)$. Then the last operator $E_3(\{e,f\})$ $\rightarrow
	g$ is processed to retrieve the results for the query.
\end{myexp}

\subsection{Physical Operators}
\label{sec:def_op}

To efficiently execute logical operators on large graphs, in this section, we
summarize 6 physical operators for them. Each of these operators could be implemented with lightweight index or in absence of index.



Note
that when one label exists multiple times in a RE, during query
processing, the two existences are treated as two labels and the
intermediate results are maintained independently to avoid the
confusion. For example,for the RE $cbc^+$, the first label $c$ and
the last label $c$ in the closure may refer to different node set
during query processing. Hence they should be distinguished.

Intuitively, the basic physical operators load nodes from the graph according the label and loading the neighbours with some specific labels. Two operators, \textsf{Load} and \textsf{Neighbour}, are defined for them, respectively.

We also propose two physical operators to link the intermediate results. One links the nodes matching some labels to intermediate results (\textsf{SingleLink}). The other joins of two groups of intermediate results (\textsf{DoubleLink}).

Corresponding to these two logical operators related closure, two special physical operators, \textsf{ClosureLink} and \textsf{SelfLink}, are required as the fixed point of finding the repeated path matching the closure.

According to above discussions, physical operators are summarized in Table~\ref{tab:phy}, where each variable may
correspond to a set of labels, as is similar as logical operators.
For example, for an RE $E$=$(a|bc|def))$, head($E$)=\{a, b, d\},
tail($E$)=\{a, c, f\}. To process the query $(h|r)(a|bc|def)$,
supposing that $E$ is processed first, the following logical
operator is
\{h,r\}$\rightarrow E(\{a,b,d\})$ corresponding physical operator list is
Load(\{h,r\}); Neighbor(\{h,r\}, \{a, b, d\}); SingleLink(\{h,r\},
\{a, b, d\},
\{a, c, f\}).

For \textsf{SingleLink}, it is supposed that the links between $a$ node and
$b$ node as well as the links between $b$ nodes and $c$ nodes have
been built. For \textsf{DoubleLink}, it is supposed that the links between
$a$ nodes and $b$ nodes, $b$ nodes and $c$ nodes , $c$ nodes and $d$
nodes have been built. For \textsf{ClosureLink}, it is supposed the links
between $a$ nodes and $b$ nodes have been built.

\begin{table*}
	\caption{Physical Operators}
	\label{tab:phy}
	\centering
	\begin{tabular}{r|p{14cm}}
		\hline
		Operator & Meaning\\
		\hline
		Load($a$) & Load the nodes with label $a$ \\
		Neighbor($a$, $b$) & Obtain $b$ neighbors of each $a$ nodes and link each $a$ node with its $b$ neighbor\\
		SingleLink($a$, $b$, $c$) & Link the $c$ neighbors of each node with
		label $b$ to all its $a$ neighbors.\\
		DoubleLink($a$, $b$, $c$, $d$) & Link each pair of $a$ and $b$ nodes
		where each $a$ node, is a neighbor of a $b$ node $N_b$, each $d$
		node is a neighbor of a $c$ node $N_c$ and $N_b$ links $N_c$ node\\
		ClosureLink($a$, $b$) & Link each $a$ node $N_a$ with all its $b$
		descendants with each such $N_b$ exits a path
		$N_aN_{b_1}N_{a_1}\cdots N_{b_n}N_{a_n}N_b$ between them, where
		$N_{a_1}$, $\cdots$, $N_{a_n}$ have label $a$; $N_{b_1}$, $\cdots$,
		$N_{b_n}$ have label $b$
		\\
		SelfLink($a$, $a_1$, $a_2$)& generate a copy of the results of $a$
		and link the nodes in two sets referring to the same nodes in the
		graph.\\
		\hline
	\end{tabular}
\end{table*}

The relationships between logical operators and physical operators
are shown in Table~\ref{tag:corres}. In Table~\ref{tag:corres}, for
closure, the function \textsf{Link} depends on the form of $E_1$ and $E_2$. If
both $E_1$ and $E_2$ are single labels, this link operator is
unnecessary; if only one of $E_1$ and $E_2$ is a complex regular
expression and the other is a single label, the operator is
\textsf{SingleLink}; if both of $E_1$ and $E_2$ are a complex regular
expressions, corresponding operator is \textsf{DoubleLink}.

\begin{table}
	\caption{The Relationships between Logical operators and Physical Operators}
	\label{tag:corres}
	\scriptsize
	\begin{tabular}{r|l}
		\hline
		Logical Operator & Physical Operator List\\
		\hline
		$a \rightarrow b$& Load($a$); Neighbor($a$, $b$)\\
		$a \leftarrow b$& Load($b$); Neighbor($b$, $a$)\\
		$a \rightarrow E(b)$ & Load($a$); Neighbor($a$, $b$); SingleLink($a$, $b$, tail($E$))\\
		$a \leftarrow E(b)$ & Neighbor($b$, $a$); SingleLink($a$, $b$, tail($E$))\\
		$E(b) \rightarrow a$ & Neighbor($b$, $a$);SingleLink(head($E$), $b$, $a$) \\
		$E(b) \leftarrow a$ & Load($a$); Neighbor($a$, $b$));SingleLink(head($E$), $b$, $a$)\\
		$E_1(a) \rightarrow E_2(b)$ & Neighbor($a$),($b$));DoubleLink(head($E_1$), $a$, $b$, tail($E_2$))\\
		\hline
		$E(E_1(a)-E_2(b))^+$ & ClosureLink($E$);Link(head($E_1$), $a$, $b$, tail($E_2$))\\
		$a^+$ & Load($a$), SelfLink($a$, $a_1$, $a_2$);ClosureLink($a_1$,$a_2$)\\
		\hline
	\end{tabular}
\end{table}

We use an example to illustrate the intermediate results and the
physical operator execution.

\begin{figure}[t]
	\subfigure[Results of $L_1$]{
		\label{fig:phy1}
		\includegraphics[width=0.3\linewidth]{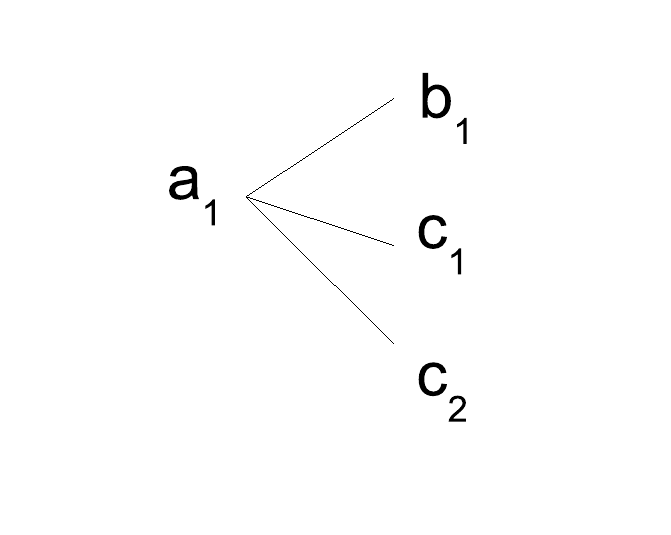}
	}
	\subfigure[Result of $P_{21}$]{
		\label{fig:phy2}
		\includegraphics[width=0.3\linewidth]{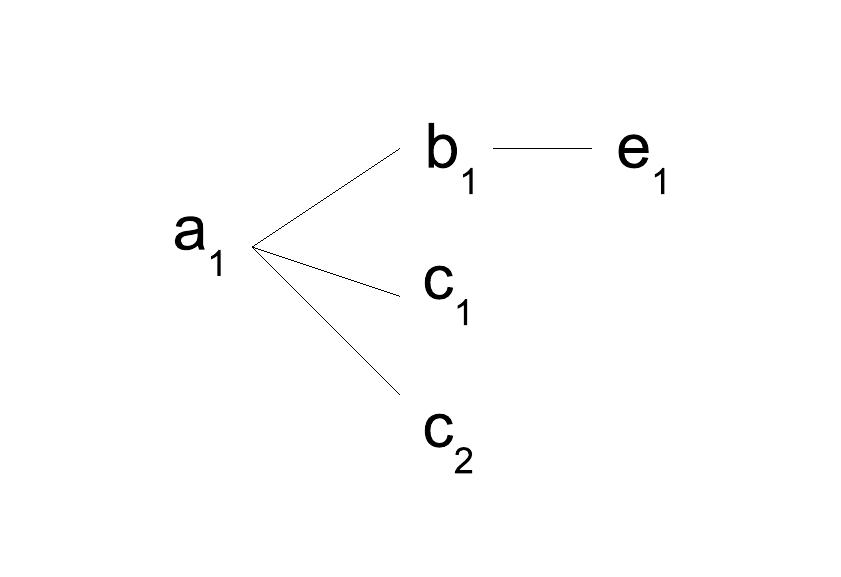}
	}
	\subfigure[Result of $P_{22}$]{
		\label{fig:phy21}
		\includegraphics[width=0.2\linewidth]{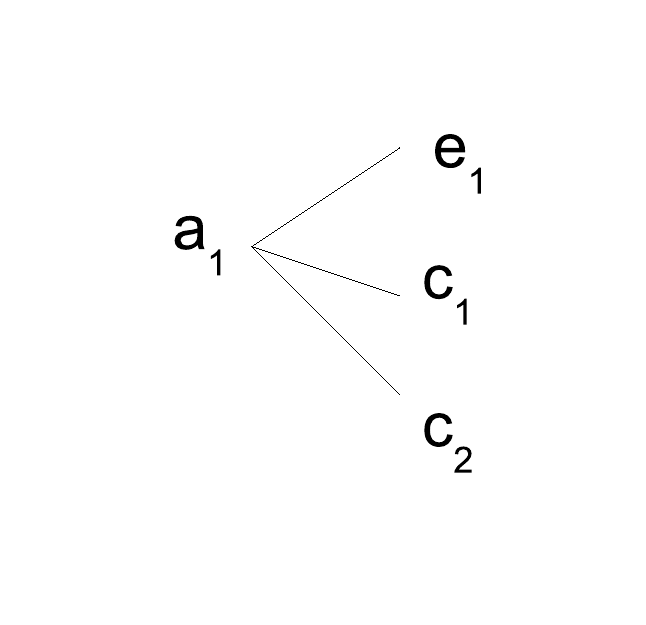}
	}
	\subfigure[Result of $P_{31}$]{
		\label{fig:phy31}
		\includegraphics[width=0.3\linewidth]{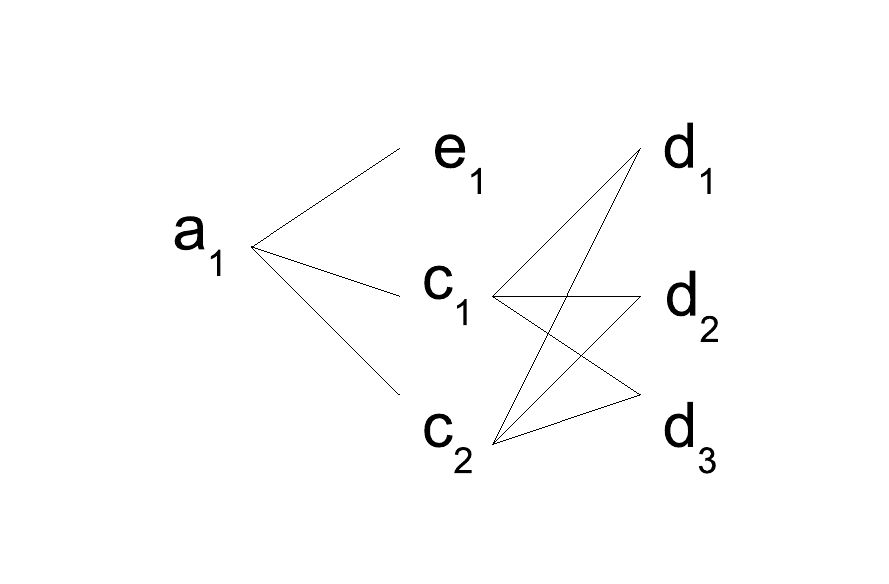}
	}
	\subfigure[Result of $P_{32}$]{
		\label{fig:phy32}
		\includegraphics[width=0.2\linewidth]{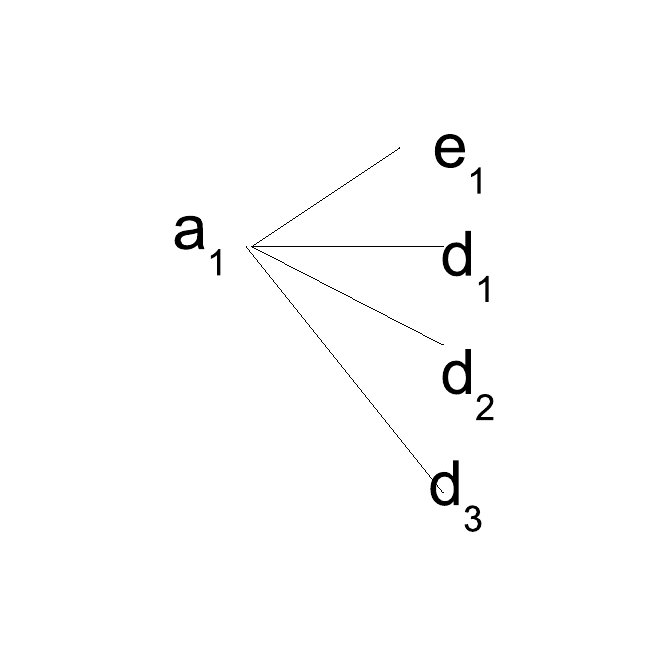}
	}
	\subfigure[Result of $P_{41}$]{
		\label{fig:phy41}
		\includegraphics[width=0.3\linewidth]{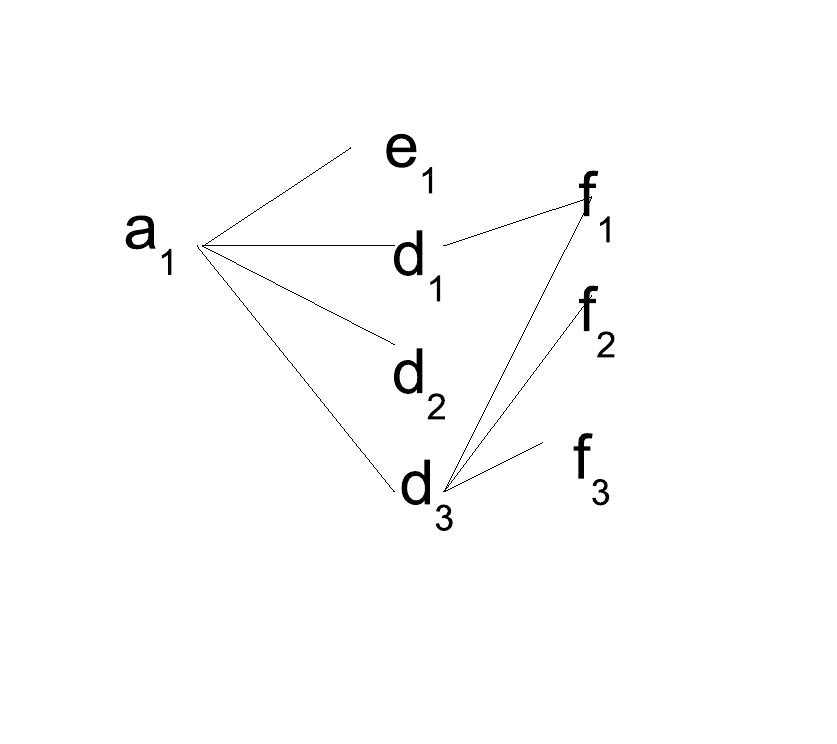}
	}
	\subfigure[Result of $P_{42}$]{
		\label{fig:phy42}
		\includegraphics[width=0.2\linewidth]{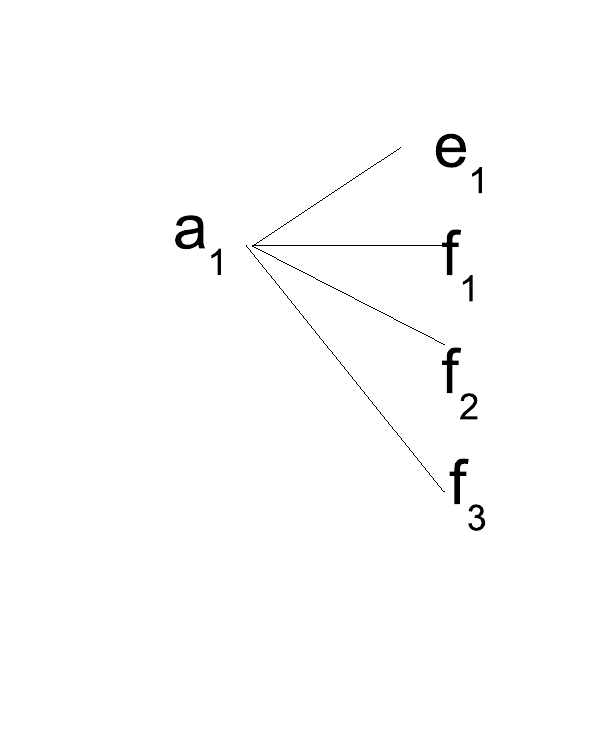}
	}
	\subfigure[Result of $P_{51}$]{
		\label{fig:phy51}
		\includegraphics[width=0.3\linewidth]{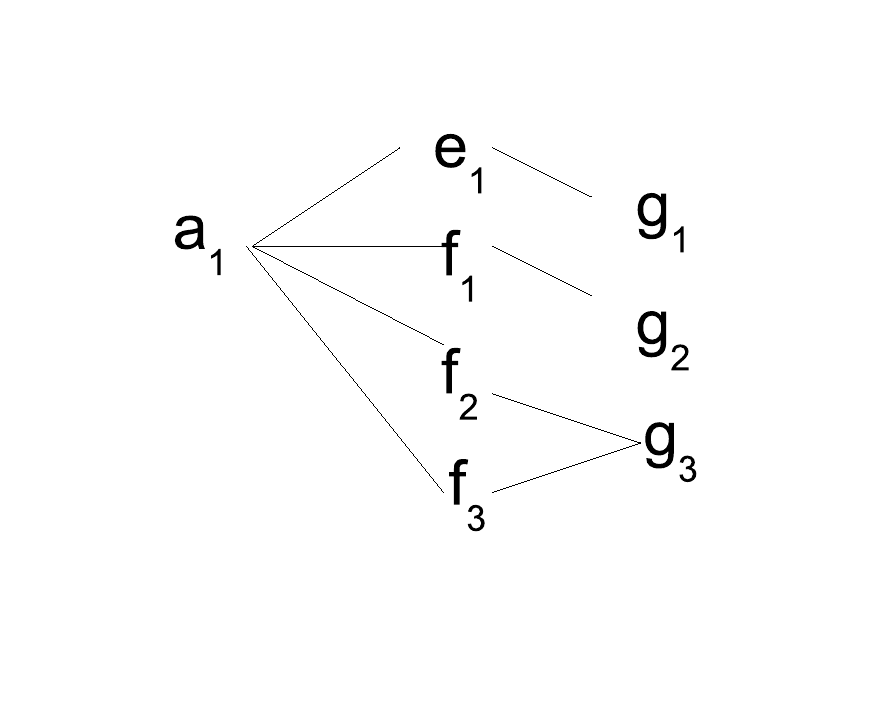}
	}
	\subfigure[Result of $P_{52}$]{
		\label{fig:phy52}
		\includegraphics[width=0.3\linewidth]{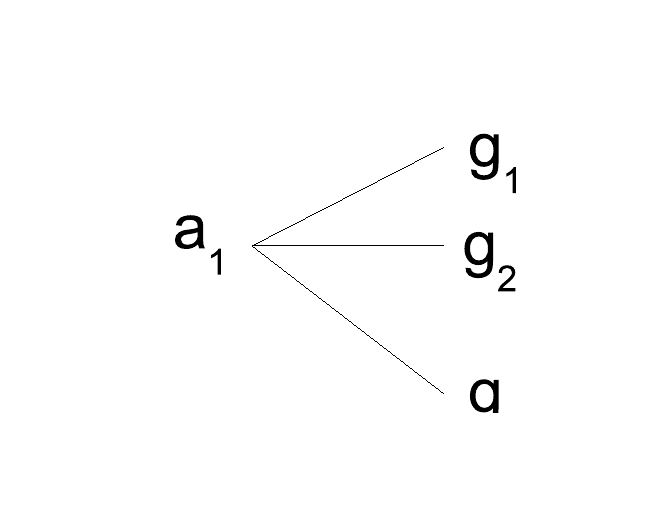}
	}
	\caption{Intermediate Results}
	\label{fig:intermedia}
\end{figure}

\begin{table}[t]
	\caption{The Logical Operators and Physical Operators for the Example}
	\label{tab:exp_corr}
	\begin{tabular}{r|l|r|l}
		\hline
		LID & Logical operator & PID & Physical Operator\\
		\hline
		$L_1$ & $a\rightarrow (b|c)(\{b,c\})$ & $P_{11}$ & Load($a$)\\
		& & $P_{12}$ & Neighbor($a$, \{$b,c$\})\\
		\hline
		$L_2$ & $a(b|c)\{b\}\rightarrow e$ & $P_{21}$ & Neighbor($b$, $e$)\\
		& & $P_{22}$ & SingleLink($a$, $b$, $e$)\\
		\hline
		$L_3$ & $(c\rightarrow d)^+(E_1(c),d)$ & $P_{31}$ & ClosureLink($c$, $d$)\\
		& & $P_{32}$ & SingleLink($a$, $c$, $d$))\\
		\hline
		$L_4$ & $E_2(d)\rightarrow f$ & $P_{41}$ & Neighbor($d$,$f$))\\
		&  & $P_{42}$ & SingleLink($a$, $d$, $f$)\\
		\hline
		$L_5$ & $E_3(\{e,f\})\rightarrow g$ & $P_{51}$ & Neighbor($\{e,f\}$, $g$)\\
		& & $P_{51}$ & SingleLink($a$, $\{e,f\}$, $g$)\\
		\hline
	\end{tabular}
\end{table}

\begin{myexp}
	\label{exp:physical}
	
	In this example, we show the execution of physical operators for the
	query and logical operator execution order in
	Example~\ref{exp:framework}. The corresponding relationships between
	logical operators and physical operators are shown in
	Table~\ref{tab:exp_corr}, where the column LID is the id of logical
	operators and PID is the id of physical operators; the meanings of
	$E_1$, $E_2$ and $E_3$ are the same as those in Example~\ref{exp:framework}.
	
	The intermediate results for the operators are shown in
	Figure~\ref{fig:intermedia}. For the second logical operator
	$a(b|c)(\{b\})\rightarrow e$, at first, the $e$ neighbor of $b$
	nodes are obtained and then \textsf{SingleLink} operator is performed to
	connect $a$ node with $e$ nodes. For the logical operator
	$(c\rightarrow d)^+(E_1(c),d)$, the first physical operator is
	ClosureLink($c$, $d$), the $c$ nodes as the input include $c_1$ and
	$c_2$, but $c_3$ is not the input of this operator. The results of
	the physical operator are ($c_1$,$d_1$), ($c_1$,$d_2$),
	($c_1$,$d_3$), ($c_2$,$d_1$), ($c_2$,$d_2$) and ($c_2$,$d_3$). Then
	$a_1$ is connected with $d_1$, $d_2$ and $d_3$.
\end{myexp}

The efficient implementation algorithms of these physical operators will be discussed in the next section.

\subsection{Logical Operator Generation}
\label{sec:LoOp}
We explain the conversion from a regular expression to logical operators. Such generation is implemented using
postfix expression.

Any infix regular expression can be easily rewritten into a postfix
expression. For example, given infix expression $a b- ( (b-c) | ( (d|e)-
f ) ) -(g|h)$, its postfix expression is $a b c - d e | f - | - g h |
-$. Given a postfix expression, we can generate its logical
operators using a tri-column stack in the schema of (operand, prefix, postfix). The operand is the operator, while prefix/postfix column stores the prefix/postfix
character set of current stack frame. The algorithm yields a logical
operator when meeting concatenation operator ``-'', concatenating the
postfix of the first operand and the prefix of the second operand. It
merges prefixes and postfixes when meeting ``or'' operator ``$|$'',
merging the prefixes/postfixes of two operands.


The logical operator generation algorithm consists of three steps:
\begin{list}{\labelitemi}{\leftmargin=1em}\itemsep 0pt \parskip 0pt
	\item Scan the postfix expression from left to right;
	\item If current character is an operand, then push it to the stack;
	\item If current character is an operator, pop two operands, doing calculation and push the result back.
\end{list}


\section{Implementation Algorithms of Physical Operators}
\label{sec:operator}
In this section, we will discuss the implementation algorithms of
physical operators in Section~\ref{sec:def_op}. For the
convenience of discussions, we introduce centralized algorithms and
distributed algorithms in Section~\ref{sec:centralized} and
Section~\ref{sec:distributed}, respectively.

\subsection{Centralized Algorithms}
\label{sec:centralized}
This section proposes the implementation algorithms for physical
operators in centralized environment. Note that the algorithms are implemented with a lightweight index or in absence of index. The description of the algorithms in this section could be easily implemented on a database with join operators.

To simplify the algorithm descriptions, the symbols and functions used
in the algorithm descriptions are shown in Table~\ref{tab:sym}.

\begin{table}
	\caption{Symbols in Physical Operator Implementation Algorithms}
	\label{tab:sym}
	\begin{tabular}{r|p{5.5cm}}
		\hline
		Symbol & Meaning \\
		\hline
		$n$ & the number of vertices in the graph\\
		$m$ & the number of edges in the graph\\
		S($a$) & label set corresponding to the variable $a$ in the operator\\
		link($n_1$, $n_2$, $r_1$, $r_2$) & add $n_1$ as the $r_1$ neighbor
		of $n_2$ and $n_2$ as the $r_2$ neighbor of $n_1$\\
		Filter($l$, $S$) & remove the nodes in $R_l$ without any link to the
		node corresponding to any symbol in $S$\\
		label($k$) & The label of the node $k$\\
		\hline
	\end{tabular}
\end{table}

Since the implementation of SelfLink is straightforward, we focus on other operators.

\noindent \underline{Load} The implementation of the physical operator Load is simply load the
nodes with label in S($a$) from the graph. With the index
that retrieve the nodes according to the label, this operator is
performed in time linear to the node number. Such index is called \emph{label index}, as an inverted list that retrieves the id sets according to the given label, and the size is O($n$).

\noindent \underline{Neighbor} The operator of Neighbor is implemented by load each neighbor with
the label in S($b$) for each node in $R_l$ ($l\in S(a)$) and link
them. At last, the nodes with label in $R_l$ without any neighbor
with node in S($b$) are filtered. For the efficiency issue, the implementation of this operator requires accessing the neighbors of corresponding nodes and merge their ids with the id set in corresponding entry in the label index or corresponding intermediate node sets to avoid accessing the labels of nodes. 


The implementation of the operator Neighbor is shown in
Algorithm~\ref{alg:neighbor}.


\begin{algorithm}[ht]
	\caption{Neighbor($a$, $b$)}
	\label{alg:neighbor}
	\begin{algorithmic}
		\FOR{each $r_1$ in S($a$)}
		\FOR{each $n_1$ in $R_{r_1}$}
		\FOR{each $r_2$ in S($b$)}
		\STATE load all neighbors of $n_1$ with label $r_1$ to $N_{n_1, r_1}$
		\FOR{each node $n_2$ in $N_{n_1, r_1}$}
		\STATE link($n_1$, $n_2$, $r_1$, $r_2$)
		\ENDFOR
		\STATE $R_{r_1}$=$R_{r_1}\bigcup N_{n_1, r_1}$
		\ENDFOR
		\ENDFOR
		\ENDFOR
		\FOR{each $r_2$ in S($b$)}
		\STATE filter($r_2$, S($a$))
		\ENDFOR
	\end{algorithmic}
\end{algorithm}

\noindent \underline{SingleLink and DoubleLink} As described in Table~\ref{tab:phy}, the goal of SingleLink and
DoubleLink is to link nodes in the head and tail in a path. In order
to reduce redundancy computation. We apply BFS strategy. As shown in Algorithm~\ref{alg:singlelink},
at first the nodes in $R_{r_2}$ ($r_2\in S(b)$)
linked by nodes in $R_{r_1}$ ($r_1\in S(a)$) are collected in $M$. (Line~\ref{line:single1}-Line~\ref{line:single2})
Then for each node $n'$ in $M$, the nodes with nodes in $R_{r_3}$
($r_3\in S(c)$) linked by $n'$ (Line~\ref{line:single3}-Line~\ref{line:single4}). At last, the nodes in each $R_{r_1}$
and each $R_{r_3}$ are filtered based on their links (Line~\ref{line:single5} and Line~\ref{line:single6}). As shown in Algorithm~\ref{alg:doublelink}, the
implementation of DoubleLink is similar as SingleLink but with an
additional step of collection neighbors in $R_{r_3}$ ($r_3\in S(c)$)
for each node in $M$ (Line~\ref{line:double1}-Line~\ref{line:double2}).

\begin{algorithm}[ht]
	\caption{SingleLink($a$, $b$, $c$)}
	\label{alg:singlelink}
	\begin{algorithmic}[1]
		\FOR{each $r_1$ in S($a$)}
		\FOR{each $n_1$ in $R_{r_1}$}
		\STATE $M$=$\varnothing$\label{line:single1}
		\FOR{each $r_2$ in S($b$)}
		\STATE $M$=$M\cup N_{n_1, r_1}$
		\ENDFOR\label{line:single2}
		\FOR{each $n'$ in $M$}\label{line:single3}
		\FOR{each $r_3$ in S($c$)}
		\FOR{each $n_2$ in $N_{n', r_3}$}
		\STATE link($n_1$, $n_2$, $r_1$, $r_3$)
		\ENDFOR
		\ENDFOR
		\ENDFOR\label{line:single4}
		\ENDFOR
		\STATE Filter($r_1$, S($c$))\label{line:single5}
		\ENDFOR
		\FOR{each $r$ in S($c$)}
		\STATE Filter($r$, S($a$))\label{line:single6}
		\ENDFOR
	\end{algorithmic}
\end{algorithm}

For example, to perform the operator SingleLink($a$, \{$e$,$f$\},
$g$) on the result in Figure~\ref{fig:phy51}, at first, $e$ and $f$ nodes connected
to $a$ node are obtained. The set of $e$ and $f$ nodes connected to
$a_1$ is $S_1$=\{$e_1$, $f_1$, $f_2$, $f_3$\}. As the result, the $g$ nodes as
the neighbors in the nodes in $S_1$ are obtained. Then, the set
\{$g_1$,$g_2$,$g_3$\} is obtained and $a_1$ is linked with
$g_1$,$g_2$ and $g_3$.

\begin{algorithm}[ht]
	\caption{DoubleLink($a$, $b$, $c$, $d$)}
	\label{alg:doublelink}
	\begin{algorithmic}[1]
		\FOR{each $r_1$ in S($a$)}
		\FOR{each $n_1$ in $R_{r_1}$}
		\STATE $M$=$\O$
		\FOR{each $r_2$ in S($b$)}
		\STATE $M$=$M\cup N_{n_1, r_2}$
		\ENDFOR
		\STATE $H$=$\O$
		\FOR{each $n'$ in $M$}\label{line:double1}
		\FOR{each $r_3$ in S($c$)}
		\STATE $H$=$H\cup N_{n', r_3}$
		\ENDFOR
		\ENDFOR\label{line:double2}
		\FOR{each $n''$ in $H$}
		\FOR{each $r_3$ in S($d$)}
		\FOR{each $n_2$ in $N_{n', r_3}$}
		\STATE link($n_1$, $n_2$, $r_1$, $r_3$)
		\ENDFOR
		\ENDFOR
		\ENDFOR
		\ENDFOR
		\STATE Filter($r_1$, S($d$))
		\ENDFOR
		\FOR{each $r$ in S($d$)}
		\STATE Filter($r$, S($a$))
		\ENDFOR
	\end{algorithmic}
\end{algorithm}

\noindent \underline{ClosureLink} The pseudo code of the implementation algorithm of ClosureLink is shown in Algorithm~\ref{alg:closureLink}. The idea is to obtain the nodes to connect with $ClosureSearch$. During the search, to avoid repeated search, we use a hash table $H$ to maintain $a$ with all descendants searched. For each vertex $n$ matching $a$, a \emph{descendant} set $M_n$ is generated for all corresponding $b$ vertices by $ClosureSearch$.

To avoid redundant search, $ClosureSearch$ is implemented in the combination of DFS and BFS. For each vertex $n$ with label $a$, BFS is applied to collect all its $b$ neighbors in $H_b$ and add all the neighbors to the $a$ ancestors of $n$ in $S$ (Line~\ref{line:closure2}-Line~\ref{line:closure3}). Then, for each node $n'$ in $H_b$, DFS is applied to obtain all its $a$ ancestors and $b$ ancestors. $ClosureSearch$ is invoked for each unvisited $a$ neighbor $n''$ of $n'$ (Line~\ref{line:closure5}). If $n''$ has been visited, the corresponding $b$ nodes are copied to the ancestors (Line~\ref{line:closure6}-Line~\ref{line:closure9}).


For example, to perform the ClosureLink($c$,$d$) on the $c$ node set
\{$c_1$,$c_2$\}, $c_1$ is to be processed at first and pushed to the stack $S$. $d_1$ is added to $H_b$ and $M_{c_1}$.
Then, the traversal starts from $d_1$, and $c_3$ is obtained.
$c_3$ is pushed to $S$. The traversal starting from $c_3$ obtains
$d_2$ and $d_3$. $d_2$ and $d_3$ are added to $M_{c_1}$ and $M_{c_3}$. When the search from $c_1$  is accomplished, $c_1$ is linked to all vertices in $M_{c_1}$. After that, since $c_3$ has been visited, $c_3$ is linked to all vertices in $M_{c_3}$.

Note that descendants may have duplications. It is caused by repeated segments in the graph matching the clause in the RE. Such duplications could be avoided by adding links to sets instead of copying sets (in Line~\ref{line:closure10} and Line~\ref{line:closure11}). From the experiments, such cases seldom occur.

\begin{algorithm}[ht]
	\caption{ClosureLink($a$, $b$)}
	\label{alg:closureLink}
	\begin{algorithmic}[1]
		\STATE $H$=$\O$
		\FOR{each $l$ in L($a$)}
		\FOR{each $n$ in $R_l$}
		\IF{$n$ is not in $H$}
		\STATE $M_n$ = $\O$
		\STATE $S$ = $\O$
		\STATE ClosureSearch($n$, $l$, $a$, $b$, $M_n$, $H$, $S$)
		\ENDIF
		\STATE add $n$ to $H$
		\FOR{each $k$ in $M_n$}
		\STATE link($n$, $k$, $r$, label($k$))
		\ENDFOR
		\ENDFOR
		\ENDFOR
		\STATE ClosureSearch($n$, $l$, $a$, $b$, $M$, $H_a$, $S$)
		\STATE Push($S$, $n$)
		\STATE $H_b$=$\varnothing$
		\FOR{each $l$ in S($b$)}
		\FOR{each $j$ in $N_{n, l}$}\label{line:closure1}
		\FOR{each $k$ in $S$} \label{line:closure2}
		\IF{$j\in R_{l}$}
		\STATE add $j$ to $M_k$
		\STATE add $j$ to $H_b$
		\ENDIF
		\ENDFOR \label{line:closure3}
		\ENDFOR  \label{line:closure4}
		\FOR{each $n'$ in $H_b$}
		\FOR{each $l'$ in S($a$)}
		\STATE $TH$=LoadNeighbor($n'$, $l'$)
		\FOR{each $n''$ in $TH$}
		\IF{$n''\notin H_a$ and $n'' \notin S$}
		\STATE ClosureSearch($n''$, $l'$, $a$, $b$, $M_{n''}$, $H_a$, $S$)\label{line:closure5}
		\STATE add $n''$ to $H_a$
		\ELSIF{$n''\in H_a$}\label{line:closure6}
		\FOR{each $r$ in $S$}
		\STATE $M_r$=$M_r\cup M_{n''}$\label{line:closure10}
		\ENDFOR\label{line:closure7}
		\ELSE\label{line:closure8}
		\FOR{$r$=top($S$) to prior($n''$)}
		\STATE $M_r$=$M_r\cup M_{n''}$\label{line:closure11}
		\ENDFOR
		\ENDIF\label{line:closure9}
		\ENDFOR
		\ENDFOR
		\ENDFOR
		\ENDFOR
		\STATE POP($S$)
	\end{algorithmic}
\end{algorithm}

Note that during the operator processing, the original graph is not modified. When a vertex $n$ in the graph is obtained, a stub node for $n$ is constructed in the intermediate result. During traversal, the adding of a link between $n$ and other node $m$ is implemented by adding such link between corresponding stub nodes.

\noindent \underline{Complexity Analysis}
For a graph with $n$ vertices and $m$ edges, obviously, in the worst case, the
time complexity of operator Load and SelfLink is O($n$) since at
most each vertex is accessed only once. In the worst case, for each
vertex $v$, the operator Neighbor will access at most $n$ vertices.
Then the time complexity of $Neighbor$ is O($n^2$).

With BFS search
strategy, for each step, each dummy vertex is processed only once
and for each vertex the vertices with special labels linked to it
will be visited. Therefore, the time complexities of SingleLink and
DoubleLink are both O($n^2$).

With the hash table in the
ClosureLink, each vertex in the graph is visited at most once and
for each visited vertex, only the vertices with special labels
linked to it are to be accessed, the number of which is at most $n$.
Therefore, the time complexity of ClosureLink in the worst case is
O($n$). Since each logical operator corresponds a constant number of
physical operators and for each RE, the number of logical operators
equals to the number of concentration and closure operators $k$, the
time complexity in the worst case of RE query processing is
O($kn^2$).

\subsection{Distributed Algorithms}
\label{sec:distributed}

To handle very large graphs, it naturally adopts a distributed platform. In order to process regular expression query on large distributed graphs efficiently,
our solution provides efficient physical operator implementation algorithms on the distributed platform to minimize the communication cost.
From the aspect of graph management, the possible communication cost is caused by traversing from the vertex in one machine to those in another machine. Thus, we should select the distributed graph management platform supporting efficient traversal. 

Motivated by this, we adopt infrastructure of Microsoft Graph Engine~\footnote{http://research.microsoft.com/en-us/projects/graphengine/},  an open-source distributed in-memory graph processing engine which supports traversal efficiently. It is underpinned by a strongly-typed in-memory key-value store and a general-purpose distributed computation engine. The following two core capabilities of Graph Engine make it an ideal platform for handling distributed large graphs with complex data schema: 1) It excels at managing a massive amount of distributed in-memory objects and providing efficient random data access over the distributed data. Fast random access is the key to many graph algorithms. 2) It excels at handling big graph data with complex schema. For example, Graph Engine is serving a Microsoft knowledge graph (about 23 TB) which has thousands of entities types and billions of nodes and edges.

Thus, such platform is suitable for efficient distributed implementation of \textsf{Neighbor}, \textsf{SingleLink}, \textsf{DoubleLink} and \textsf{ClosureLink}.

In distributed environment, for the convenience of processing, each
node is stored locally, and only
the ids of nodes are transmitted during query processing.

With the consideration of network
issues, the \textsf{Load} and \textsf{SelfLink} are only performed locally. However, for other operators, the network
communications may be involved.

Clearly, a pretty graph partition among the machines in a cluster could accelerate the processing. However, the graph partition is costly in computation especially for very large graphs. Additionally, in real-world scenarios, the workload on a graph distributed in a cluster may contain various operations instead of only regular expression query processing. Thus, it is difficult to choose a proper partition criteria. To make our approach suitable for real applications, we suppose that the graph is partitioned among machines randomly without any sophisticated graph partition strategy.

\noindent \underline{Basic Operations} From the implementation strategy, the basic operations related to
two nodes are fetching a neighbor from a node and connecting two
nodes. Therefore, the queries are processed in distributed
environment by two network primitives according to the two
operations for the nodes distributed in different machines in the
network.

\begin{itemize}
	\item GetNeighbor($n$, $t_r$, $t_n$) obtains the neighbors with tag $t_r$ for the nodes $n$ with tag $t_n$.
	
	\item AddLink($u$, $v$, $a$, $b$) connects $u$ and $v$ with $u$ as $v$'s a neighbor and $v$ as $u$'s b
	neighbor.
\end{itemize}

In order to save the communication cost and network bandwidth, the
distributed implementation has two strategies. One is to load node ids
instead of all information of nodes.
The other is to load and
revise link in batch style. The batch-style operators are \textsf{BatchGetNeighbor} and \textsf{BatchAddLink}, respectively. The input of \textsf{BatchGetNeighbor} is a table in schema ($id$, $r_1$, $r_2$), with the same semantics as that of
\textsf{GetNeighbor}. The returned results are in schema
($pid$,$plabel$,$cid$,$clabel$), where $pid$ is the id of the input
tuple, $plabel$ is its label, $cid$ is the id of the obtained neighbor,
and $clabel$ is the label of the neighbor. The input of \textsf{BatchAddLink} is in schema ($pid$,$plabel$,$cid$,$clabel$), where $pid$ is the id of the input
tuple, $plabel$ is its label, $cid$ is the id of the neighbor, and $clabel$ is the label of this neighbor to link.

In the intermediate results, for each node, the links to neighbors
are the ids of the node. In each machine, to access the real node efficiently, a hash table is maintained
to map the id to the real node. And in the whole system, the id of
computer for each node $n$ is encoded in the id of $n$.

\noindent \underline{\textsf{SingleLink}, \textsf{GetNeighbour} and \textsf{DoubleLink},} The implementation of SingleLink operator share the same flow as Algorithm~\ref{alg:singlelink}. The difference is that the
functions of \textsf{BatchGetNeighbor} and \textsf{BatchAddLink} are invoked for execution in batch. The pseudo
code of the implementations of \textsf{SingleLink} operators is shown in
Algorithm~\ref{alg:netsinglelink}. Such algorithm is
executed in all machines in the system. The batch execution steps are in Line~\ref{line:dissingle1} and Line~\ref{line:dissingle2}, respectively. During the invocation of these two functions, these tasks related to the same
computer are sent to it in batch.

\begin{algorithm}[ht]
	\caption{DistributedSingleLink($a$, $b$, $c$)}
	\label{alg:netsinglelink}
	\begin{algorithmic}
		\FOR{each $r_1$ in S($a$)}
		\FOR{each $n_1$ in $R_{r_1}$)}
		\STATE $M$=$\varnothing$
		\FOR{each $r_2$ in S($b$)}
		\FOR{each $n_2$ in $N_{n_1, r_2}$}
		\STATE add ($n_1$, $r_1$, $n_2$, $r_2$) to $T_1$
		\ENDFOR
		\ENDFOR
		\ENDFOR
		\ENDFOR
		\FOR{each $t$ in $R_1$}
		\STATE add ($t[pid]$, $t[plabel]$) to $M_{t[cid],t[clabel]}$
		\FOR{each $r$ in S($c$)}
		\STATE add($t[cid]$, $t[clabel]$, $r$ to $T_2$)
		\ENDFOR
		\ENDFOR
		\STATE $R_2$=BachGetNeighbor($T_2$)\label{line:dissingle1}
		\FOR{each $t$ in $R_2$}
		\FOR{each $r$ in $M_{t[pid], t[plabel]}$}
		\STATE add ($r[id]$, $r[label]$, $t[cid]$, $t[clabel]$) to
		$T_3$
		\ENDFOR
		\ENDFOR
		\STATE BatchAddLink($T_3$)\label{line:dissingle2}
		\FOR{each $r_1$ in S($a$)}
		\STATE Filter($r_1$, S($c$))
		\ENDFOR
		\FOR{each $r_2$ in S($c$)}
		\STATE Filter($r_2$, S($a$))
		\ENDFOR
	\end{algorithmic}
\end{algorithm}

We use an example to illustrate the distributed implementation of
\textsf{SingleLink}. The distributed implementation of \textsf{DoubleLink} and
Neighbor are similar.

\begin{figure}[ht]
	\centering
	\includegraphics[width=0.6\linewidth]{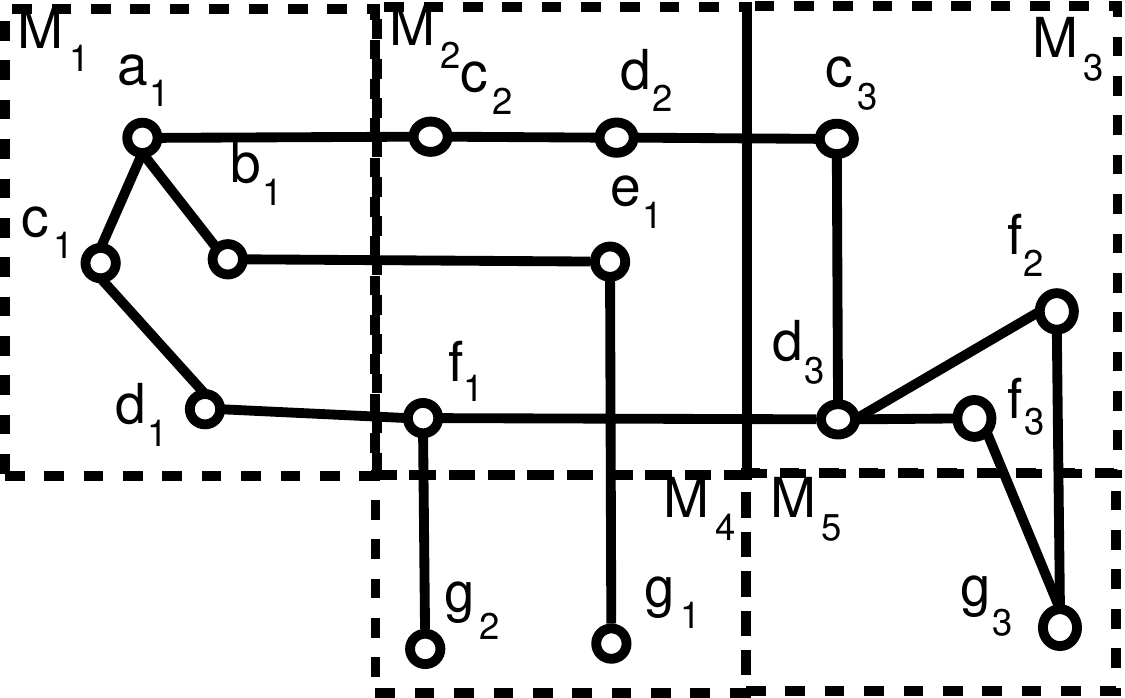}
	\caption{The distributed Graph}  \label{fig:partition}
\end{figure}

\begin{myexp}
	Consider the process of $Q_1$ in the graph in
	Figure~\ref{fig:data_graph}. It is supposed that there five computers
	$M_1$ $\cdots$ $M_5$ in the cluster.  The partition of nodes in the cluster is shown in Figure~\ref{fig:partition}.
	For the execution
	physical operator SingleLink($a$, \{$e$,$f$\}, $g$) on the
	intermediate results shown in Figure~\ref{fig:intermedia}, at first, starting from $a_1$,
	the network primitives of GetNeighbor($a_1$,
	\{$e_1$,$f_1$\}, $g$) and GetNeighbor($a_1$,
	\{$f_2$,$f_3$\}, $g$) are generated. Then, they are sent to $M_2$ and
	$M_3$, respectively. After $M_2$ and $M_3$ executed the primitives.
	The ids of $g_1$, $g_2$ and $g_3$ are returned to $M_1$. $a_1$ is
	linked with $g_1$, $g_2$ and $g_3$. With these ids, the locations of
	$g_1$, $g_2$ and $g_3$ are known. With these locations, the
	primitives of link, AddLink($a_1$,
	\{$g_1$, $g_2$\}, $a$, $g$) and  AddLink($a_1$, $g_3$, $a$, $g$),
	are generated and sent to $M_4$ and $M_5$, respectively. With these
	primitives, in $M_4$, $g_1$ and $g_2$ are linked with $a_1$, and in
	$M_5$, $g_3$ is linked with $a_1$.
\end{myexp}

\noindent \underline{ClosureLink} DFS in Algorithm~\ref{alg:closureLink} may access a remote machine multiple times for each round of iteration. However, since multiple remote accessing involve large communication cost, the implementation of \textsf{ClosureLink} on a distributed graph tries
to avoid DFS. Instead, as shown in Line~\ref{line:disclosure1} and Line~\ref{line:disclosure2} in Algorithm~\ref{alg:netclosure}, $a$ and $b$ nodes are obtained in batch. Then, Line~\ref{line:disclosure3} adds corresponding links in the intermediate results in batch.

\begin{algorithm}[ht]
	\caption{DistributedClosureLink($a$, $b$)}
	\label{alg:netclosure}
	\begin{algorithmic}[1]
		\STATE $H$=$\varnothing$
		\FOR{each $l$ in S($a$)}
		\FOR{each $n$ in $R_l$}
		\FOR{each $l'$ in S($b$)}
		\IF{($n$, $l$, $l'$) is not in $H$}
		\STATE add ($n$, $l$, $l'$) to $H$
		\FOR{each $n'$ in $N_{n,l'}$}
		\STATE add ($n$, $l$, $n'$, $l'$) to $R_1$
		\ENDFOR
		\ENDIF
		\ENDFOR
		\ENDFOR
		\ENDFOR
		\WHILE{$R_1\neq \varnothing$}
		\STATE $N$=$\varnothing$
		\FOR{each $t$ in $R_1$}
		\STATE add ($t[plabel]$, $t[pid]$) to $M_1[(t[clabel], t[cid])]$
		\STATE ($t[clabel]$, $t[cid]$) to $N$
		\ENDFOR
		\FOR{each $t$ in $N$}
		\FOR{each $l$ in S($a$)}
		\STATE add ($t[cid]$, $t[clabel]$, Label($l$)) to $T_2$
		\ENDFOR
		\ENDFOR
		\STATE $R_2$ = BatchGetNeighbor($T_2$)\label{line:disclosure1}
		\STATE $N$=$\varnothing$
		\FOR{each $t$ in $R_2$}
		\STATE add $M_1(t[plabel], t[pid]$) to $M_2[(t[clabel], t[cid])]$
		\STATE ($t[clabel]$, $t[cid]$) to $N$
		\ENDFOR
		\FOR{each $t$ in $N$}
		\FOR{each $l'$ in L($b$)}
		\IF{($n$, $l$, $l'$) is not in $H$}
		\STATE add ($n$, $l$, $l'$) to $H$
		\STATE add ($n$, $l$, $l'$) to $T_1$
		\ENDIF
		\ENDFOR
		\ENDFOR
		\STATE $R_1$=BatchGetNeighbor($T_1$)\label{line:disclosure2}
		\FOR{each $t$ in $R_1$}
		\FOR{each $r$ in $M_2[(t[plabel], t[pid])]$}
		\STATE add ($r[id]$, $r[label]$, $t[cid]$, $t[clabel]$)
		to $T_3$
		\ENDFOR
		\ENDFOR
		\STATE BatchAddLink($T_3$)\label{line:disclosure3}
		\ENDWHILE
		\FOR{each $r_1$ in S($a$)}
		\STATE Filter($r_1$, S($b$))
		\ENDFOR
		\FOR{each $r_2$ in S($b$)}
		\STATE Filter($r_2$, S($a$))
		\ENDFOR
	\end{algorithmic}
\end{algorithm}

Example~\ref{exp:phy_closure} is used to show that processing of the
distributed implementation of ClosureLink.

\begin{myexp}
	\label{exp:phy_closure}
	To process ClosureLink($c$, $d$) on the graph in Figure~\ref{fig:data_graph} with graph partition in Figure~\ref{fig:partition}.
	At first,
	the search starts from $c_1$ and $c_2$ in $M_1$ and $M_2$,
	respectively. As the result, $d_1$ and $d_2$ are obtained. The
	primitives GetNeighbor($d_1$, $c$, $d$) and GetNeighbor($d_2$,
	$c$, $d$) then request $c$ nodes from $M_3$. They are processed in $M_3$ in batch. From the
	information stored in $c_3$, the ids of $d_1$, $d_2$ and $d_3$ are
	sent to $M_1$ and $M_2$, respectively.
	
	In $M_1$, $c_1$ is linked with $d_2$ and
	$d_3$, and the primitives AddLink($c_1$,$d_2$, $c$, $d$) and
	AddLink($c_1$, $d_3$, $c$, $d$) are generated and sent to $M_2$
	and $M_3$, respectively. In $M_2$, $c_2$ is linked with $d_1$ and
	$d_2$, and the primitive AddLink($c_2$,$d_1$, $c$, $d$) and
	AddLink($c_2$, $d_3$, $c$, $d$) is generated. They are sent to $M_1$
	and $M_2$, respectively.
	
	Then, in $M_1$, $d_1$ is connected with
	$c_3$; in $M_2$, $d_2$ is linked with $c_1$; in $M_3$, $c_1$ and
	$c_2$ are linked with $d_3$. From the aspect of $M_1$, the
	search then starts from $d_2$ with $c_1d_1c_3d_2$ as the traversal path. $c_2$ is obtained. Since $c_2$ has no other $d$
	neighbor. The traversal in $M_1$ halts. Similar process is performed
	in $M_2$.
\end{myexp}

\section{Query Optimization}
\label{sec:opt}
As discussed in Section~\ref{sec:framework}, each RE query may have
different processing orders and directions of the operators. The
execution order may affect the efficiency of query processing. For
example, even for the simple query $abc$, it is supposed that each of 1G
nodes with tag $a$ is linked with one different $b$ node and
only one of the $b$ nodes is connected with a $c$ node. If the query
is executed in the direction from $a$ to $c$, more
than 2G nodes are accessed during search, while only
two nodes are accessed if the query is executed from $c$ to $a$.

Motivated by this, in this section, we develop the query optimization
strategy for RE queries. As the base of query optimization, we propose the cost estimation methods for operators and the
size of intermediate results in Section~\ref{sec:esti}.

\subsection{Estimation}
\label{sec:esti}
In this section, we propose a cost-based query estimation method
for query optimization. Since for a large graph, it is impossible to
maintain a global sketch with size super linear to the graph size. Therefore, in our system, only following
statistic information is kept, which is independent to the graph size.

\begin{enumerate}
	\item The number of the nodes with label $l$, denoted by $Num(l)$
	
	\item The average neighbor number of the nodes with label
	$l$, denoted by $Neighbor(l)$.
	
	\item The probability of a node with label $l_1$ has at least one neighbor
	with label $l_2$ respectively, denoted by $Pro(l_1, l_2)$.
	
	\item The average number of the neighbor with label $l_2$ of a node with
	label
	$l_1$ if it has at least one neighbor with $l_2$, denoted by
	$T_{Neighbor}(l_1, l_2)$
\end{enumerate}

Such information is computed by traversing the graph once. In a
large graph, the size of $Num(l)$ and $Neighbor(l)$ for all labels
is linear to the size of the label number $L$. However, the size of
of $Pro$ and $T_Neighbor$ is $L^2$. In order to deal with the graph
with a large number of labels, we use two thresholds $\epsilon_P$ and
$\epsilon_T$. For two labels $l_1$ and $l_2$, if $Pro(l_1,
l_2)<\epsilon_T$, a small value $\delta_p$ is used instead of the
real value of $Pro(l_1, l_2)$. Similarly,  if $T_Neighbor(l_1,
l_2)<\epsilon_T$, the real value of $T_{Neighbor(l_1, l_2)}$ is
replaced by a small value $\delta_T$ during optimization. Thus, the small size of statistic information is ensured.

Based on above
statistics information, we introduce the estimation approach. As discussed in Section~\ref{sec:operator},
one label may exist multiple times in a RE corresponding to
different set of intermediate results. During query
processing, the size of intermediate results will change. In order
to distinguish multiple occurrences of the same label in the
optimization, we assign a uniform id for the occurrences of each label and use L($x$) to denote the corresponding label of
the label with id $x$. We use $I_{x}$ to denote the set of
intermediate results corresponding to $x$. In the remaining part of
this section, without explicitly explanation, the variables in the
operators and formulas refer to a set of ids instead of labels.

For the convenience of discussions, we discuss the estimation of operators without '$|$' as the basic version. Then we discuss the extension to the operators with '$|$'

\subsubsection{The Estimation of Operators without '$|$'}

For
the brief of discussion, in the beginning,
we focus on the estimation for operators without `$|$'.
The subqueries with `$|$' will be discussed later. It means that
each variable refers to single id.

\begin{table}
	\caption{Symbols in Estimation}
	\label{tab:sym}
	\centering
	\begin{tabular}{r|p{6.5cm}}
		\hline
		Symbol & Meaning\\
		\hline
		$Link'(a,b)$ & the average number nodes in $I_b$ linked by each
		node in $I_a$ after the operator computation\\
		$Link(a,b)$ & the average number nodes in $I_b$ linked by each
		node in $I_a$ before the operator computation\\
		$size'(a)$ & the size of $I_a$ before the operator computation\\
		$size(a)$ & the size of $I_a$ after the operator computation\\
		\hline
	\end{tabular}
\end{table}

\noindent \underline{Basic Information} The cost estimation is based on the sizes of intermediate results and
the number of links between the nodes in two intermediate result
sets. Thus, we discuss the estimation of such parameters at first.
The symbols used in such estimation are described in Table~\ref{tab:sym}. Then we discuss the size and link number estimation in the following paragraphs. The results are summarized in Table~\ref{tab:size}, where the
column Link is the number of links between of two sets of
intermediate results, and column Size is the numbers of the intermediate
results.

\noindent \underline{Result Size of \textsf{SingleLink} and \textsf{DoubleLink}} The estimated size of \textsf{Load} and \textsf{SelfLink} is intuitively the number of corresponding nodes. We
use \textsf{SingleLink} as an example to explain the idea of the estimation \textsf{SingleLink} and
\textsf{DoubleLink}. For the operator \textsf{SingleLink}, Link($a$,$b$)Pro($b$,$c$)
is the average number of links between $I_a$ and the nodes in $I_b$
with at least one $c$ node. With Link($b$,$c$) as the average
number of links between $I_b$ and $I_c$, for each node in $I_b$, the estimated link number between $I_a$ and $I_b$ is
multiplied by Link($b$,$c$) to compute the average number of links between
each node in $I_a$ and the nodes in $I_c$ through some nodes in $I_b$. Since
each node $n$ in $I_a$ is connected with a node in $I_b$, the
probability of $n$ links with at least a node in $I_c$ through the
node in $I_b$ is the probability that a node in $I_b$ links to at
least a node in $I_c$. The estimation of the size of $I_c$ after the
operator execution is similar.

\noindent \underline{Result Size of \textsf{ClosureLink}} In the formula of estimated
links in ClosureLink($a$, $b$), $t$ means the average
number of link between each $b$ node $n_b$ in $I_b$ and the node in
$I_b$ through the path $n_b-n_a-n_b'$, where $n_a\in I_a$ and
$n_b'\in I_b$. In $t$, Pro($b$, $a$)$T_{Neighbor}(b, a)$ is the
average number of the neighbors with label L($a$) of each node of
$I_b$ and Pro($b$, $a$)$T_{Neighbor}(b,a)$ is the average number of
nodes with label L($a$) of each node of $I_b$. Pro($a$,
$b$)$link'(a,b)$ is the average number of nodes with label L($b$) linked
by each node with label L($a$) in the intermediate result. Hence,
$t$ represents the average number of nodes in $I_b$ of each node in
$I_a$ with a path $a-b$ between them. Since the computation of
ClosureLink requires the traversal of the paths in label form
$a-b-a-\cdots-a-b$ with the all possible lengths, the average number
of links in the results is estimated as link'($a$, $b$)+ link'($a$,
$b$)$t$+ $\cdots$ +link'($a$, $b$)$t^k$ + $\cdots$. When $t\leq 1$, since the maximal number of links
between each node in $I_a$ and the nodes in $I_b$ is $size'(b)$, the
estimation of average node number in $b$ lined with each node in
$I_a$ is $size'(b)$. Since before \textsf{ClosureLink} operation, each node
in $I_a$ has been linked with at least one node in $I_b$, and each node in
$I_b$ has been linked with at least one node in $I_a$, the size of
$I_a$ and $I_b$ are not changed after the operation.

\begin{table*}
	\caption{The Intermediate Result Size of Operators}
	\label{tab:size}
	\centering
	\begin{tabular}{r|p{10cm}|p{4cm}}
		\hline
		Operator & Link & Size\\
		\hline
		Load($a$) & - & size($a$)=Num($a$)\\
		Neighbor($a$, $b$) & $size'(a)$Pro($a$,$b$)Neighbor(L($b$)) & size($a$)=$size'(a)$Pro($a$,$b$); size($b$)=Num($b$)Pro($b$,$a$)\\
		SingleLink($a$, $b$, $c$) & Link($a$,$b$)Pro($b$,$c$)Link($b$,$c$) & size($a$)=$size'(a)$Pro($b$,$c$); size($c$)=$size'(c)$)Pro($b$,$a$)\\
		DoubleLink($a$, $b$, $c$, $d$) & Link($a$,$b$)Pro($b$,$c$))Link($b$,$c$)Pro($c$,$d$)Link($c$,$d$) & size($a$)=$size'(a)$Pro($b$,$c$)Pro($c$,$d$); size($d$)=$size'(d)$Pro($c$,$b$)Pro($b$,$a$)\\
		ClosureLink($a$, $b$) & Link($a$,$b$)$\frac{1}{1-t}$($t<1$);
		$size'(b)$ ($t>1$); $t$=Pro($b$, $a$)$(T_{Neighbor}(b, a)-1)$Pro($a$,
		$b$)$(Link'(a, b)-1)$& size($a$)=$size'(a)$; size($b$)=$size'(b)$
		\\
		SelfLink($a$)& Num($a$) & size($a$)=Num($a$)\\
		\hline
	\end{tabular}
\end{table*}

With this information, the cost estimation formulas of physical
operators are shown in Table~\ref{tab:cost}.

\noindent \underline{Cost of \textsf{SingleLink} and \textsf{DoubleLink}} Since the operators
\textsf{Load} and \textsf{SelfLink} access each node in the candidate only once, their
costs are the same as the number of candidates. The operator \textsf{SingleLink}
has two phases, the first phase collects the nodes in $I_b$
for the nodes in $I_a$ with cost as $size'(a)$Link($a$,$b$), and
the second phase collects the nodes in $I_c$ linked with nodes in
$I_a$ with cost $size'(a)$Link($a$,$b$)Link($b$, $c$). The
estimation for \textsf{DoubleLink} is similar.

\noindent \underline{Cost of ClosureLink} As shown in
Algorithm~\ref{alg:closureLink}, the operator ClosureLink has
multiple phases, each phase has two steps. The first is to obtain
the nodes in $I_b$ that is connected with the nodes in $I_a$ which
require to access the neighbors of each node in $I_b$. From the
aspect of starting from a single node in $I_a$, in the first step,
the cost is Link'($a$,$b$). In the first step of the next phase,
the number of the starting nodes in $I_a$ is
size'($b$)pro($b$,$a$)$\frac{size'(a)}{Num(a)}$, since the number of
all nodes with label L($b$) in the second step of the last phase is
size'($b$). The cost of the first step in the second phase is
size'($b$)pro($b$,$a$)$\frac{size'(a)}{Num(a)}$Link'($a$,$b$). With
the assumption that the share of intermediate results in $I_b$ is
linear with the that in $I_a$, the number for nodes in $I_b$ in this
step is

\begin{displaymath}
	\scriptsize
	size'(b)pro(b,a)\frac{size'(a)}{Num(a)}\frac{size'(b)}{size'(a)}
	= size'(b)pro(b,a)\frac{size'(b)}{Num(a)}
\end{displaymath}

Therefore,
the number of nodes in $I_a$ as the input of the first step of the
third phase is

\begin{displaymath}
	size'(b)Pro(b,a)\frac{size'(b)}{Num(a)}Pro(b,a)\frac{size'(a)}{Num(a)}
\end{displaymath}

Therefore the cost of the first step of the $k$th phase ($k>1$) is
size'($b$)$r^{k-1}\frac{size'(b)}{size'(a)}$Link'($a$,$b$), where
$r$=Pro($b$,$a$)$\frac{size'(a)}{Num(a)}$. With the consideration
that all $b$ nodes are accessed only once, the cost of the second
step of all the phases is I($b$)$T_{Neighbor}(b,a)$. For the case that
$r$ is larger than 1, the cost is estimated as the maximum cost of
search all possible nodes with labels L($a$) and L($b$).

\begin{table}
	\caption{The Cost of Operators}
	\label{tab:cost}
	\centering
	\begin{tabular}{r|p{5.5cm}}
		\hline
		Operator & Cost \\
		\hline
		Load($a$) & Num(L($a$)) \\
		Neighbor($a$, $b$) & $size'(a)$Neighbor(L($a$)) \\
		SingleLink($a$, $b$, $c$) & $size'(a)$Link($a$,$b$)(1+Link($b$,$c$))\\
		DoubleLink($a$, $b$, $c$, $d$) & $size'(a)$Link($a$,$b$)(1+Link($b$,$c$)(1+Link($c$,$d$)))\\
		\multirow{2}{*}{ClosureLink($a$, $b$)} &
		size'(a)Link'($a$,$b$)+$\frac{(size'(b))^2}{size'(a)}$$\frac{1}{1-r}$Link'($a$,$b$) + $I$($b$)$T_{Neighbor}(b,a)$
		($r<1$)\\
		&Num($a$)$T_{Neighbor}(a)$+Num($b$)$T_{Neighbor}(b)$ ($r>1$)
		\\
		SelfLink($a$)& size($a$)\\
		\hline
	\end{tabular}
\end{table}

\subsubsection{The Estimation of Operators with `$|$'}

Above estimation focuses on the operators without `$|$' in the
input. Then we discuss the estimation for the cases with
`$|$', where each variable in the operator may refer to multiple
labels.

For the operators of Load and \textsf{SelfLink}, the cost and size is the sum the cost and result size of all labels.

Since the
operators of \textsf{Neighbor}, \textsf{SingleLink} and \textsf{DoubleLink} with variables
referring to multiple labels can be considered as the execution of a
series of operators with the same type and variables referring to
single label, their costs and link numbers can be estimated as the sum of the cost and the number of
links of all possible combination of the Cartesian production of the
candidate sets, respectively.

For example, the results of the operator SingleLink($a_1|\cdots
|a_m$, $b_1|\cdots |b_n$, $c_1|\cdots |c_k$) is considered
as a set of operators with each one SingleLink($a_i$, $b_j$, $c_l$)
where $i\in [1, m]$, $i\in [j, n]$ and $l\in [1, k]$. Therefore, its
cost is estimated as follows.
\begin{displaymath}
	\sum_{i=1}^{m}{\sum_{j=1}^{n}{Link(a_i,b_j)\sum_{l=1}^{k}{Link(b_j,c_l)}}}
\end{displaymath}

Its number of links is estimated as
\begin{displaymath}
	Link(a_i,c_l=\sum_{j=1}^{n}{Link(a_i,b_j)Pro(b_j,c_l)Link(b_j,c_l)\frac{size'(c_l)}{N_{L(c_l)}}}
\end{displaymath}

For $a_i$,
\begin{displaymath}
	size(a_i)=max\{size'(a_i),size'(a_i)\sum_{j=1}^{n}{\sum{l=1}^{k}{Pro(b_j,c_l)}}\}
\end{displaymath}

For \textsf{ClosureLink}, since the estimation involves multiple nodes, we
use a matrix to represent the estimation. For example, for the
operator ClosureLink($a|b$,$c|d$), the patterns of a possible
matched path in length 4 may be $acac$, $acad$, $acbd$, etc. The
estimation of the cost of traversing such path require the
consideration of all possible cases. To enumerate all the cases for
the formula of \textsf{ClosureLink} cost computation, we use matrices to
represent the links, probabilities and the number of neighbors between
labels and use the computation on the matrix to compute the costs
and the links. The matrices used in the estimation are listed as
follows. $A$ is an $m\times m$ matrix, in which
$A_{ii}$=$size'(a_i)$, $A_{ij}$=0 ($i\neq j$); $B$ is an $n\times n$
matrix, in which $B_{ii}$=$size'(b_i)$, $B_{ij}$=0 ($i\neq j$); $P$
is an $n\times m$ matrix with each entry $P_{ij}$=$Pro(b_i,
a_j)\frac{size'(a_j)}{I_{a_j}}$; $R$ is an $m\times n$ matrix with
each entry $R_{ij}$=$\frac{size'(b_j)}{size'(a_i)}$; $L$ is an
$m\times n$ matrix with each entry $L_{ij}$=Link'($a_i$,$b_j$).

Then
based on the discussion in the cost estimation for \textsf{ClosureLink} with
variables referring to single labels, the cost of the first step in the first phase is estimated as $F$=$AL$, where each entry $F_{ij}$
represents the cost for the searching from the nodes in $I_{a_i}$ to
the nodes in $I_{b_j}$. For the first step in the second phase, each
entry $r_{ij}$ in $BP$ represents the number of nodes in $I_{a_j}$
linked by a node from $I_{b_i}$. Thus, the cost of the first step in
the second phase is estimated as a $n\times n$ matrix $BPL$ with
each entry $r_{ij}$ in it representing the cost of linking of the
nodes in $I_{b_j}$ from all nodes from $\bigcup_{1\leq k\leq
	m}{I_{a_k}}$ linked by the nodes from $I_{b_j}$. Similarly, the
input size in the third phase is estimated as $BPTP$ and the cost is
estimated as the matrix $BPTPL$. In summary, the cost for
ClosureLink($a_1|a_2|\cdots |a_m$,$b_1|b_2|\cdots |b_n$) is
estimated as follows.
\begin{displaymath}
	\sum_{1\leq i\leq m, 1\leq j\leq
		n}{F_{ij}}+\sum_{1\leq i\leq n, 1\leq j\leq n}{E_{ij}}
\end{displaymath}

where
$F$=$AL$ and $E$=$BPL$+$BPRPL$+$\cdots$. Note that the number of
items in the computation formula of $R$ is smaller than the diameter
of the graph. To accelerate the processing, we give a constant $r$,
only the sum of the first $r$ items is used.

The link number of \textsf{ClosureLink} is estimated by phases. The
link number generated in the first phases is $L$, each entry
$L_{ij}$ of which represents the number of original link between
$I_{a_i}$ and $I_{b_j}$. Following the link number estimation
formula for ClosureLink in Table~\ref{tab:size}, the links generated
in the second phase is estimated as the matrix $LT_1T_2$ with each
entry $r_{ij}$ as the number of additional links from $I_{a_i}$ to
$I_{b_j}$ in this phase, where $T_1$ is an $n\times m$ matrix with
each entry $T_{1ij}$= Pro($b_i$, $a_j$)$(T_{Neighbor}(b_i, a_j)-1)$
and $T_2$ is an $m\times n$ matrix with each entry
$T_{2ij}$=Pro($a_i$, $b_j$)$(Link'(a_i, b_j)-1)$. Note that each
entry in $T_1T_2$ is the sum of the items in form of Pro($b_{i}$,
$a_{j}$)$(T_{Neighbor}(b_{i}, a_{j})-1)$Pro($a_{j}$,
$b_k$)$(Link'(a_j, b_k)-1)$, which is consistent with the formula of
$t$ in Table~\ref{tab:size}.

Therefore, the number of links between each pair are computed as
following formula with each $N_{ij}$ is the number of links between
each node in $I_{a_i}$ and each node in $I_{b_j}$ in the results of
ClosureLink($a_1|\cdots |a_m$, $b_1|b_2|\cdots |b_n$).
\begin{displaymath}
	N=L+LT_1T_2+LT_1T_2T_1T_2+\cdots
\end{displaymath}

Similar as the cost estimation of ClosureLink, to accelerate the
processing, we set a constant $r$, and only the sum of the first $r$
items is used.

\subsection{Query Optimization Algorithm}
To obtain the optimal query plan for a query, we develop the query
optimization algorithm for regular expression query based on the
cost model introduced in the last section. Since each logical plan
operator corresponding to a fixed series of physical operators. We
focus on logical operators in this section.

\begin{figure}
	\centering
	\includegraphics[width=0.3\textwidth]{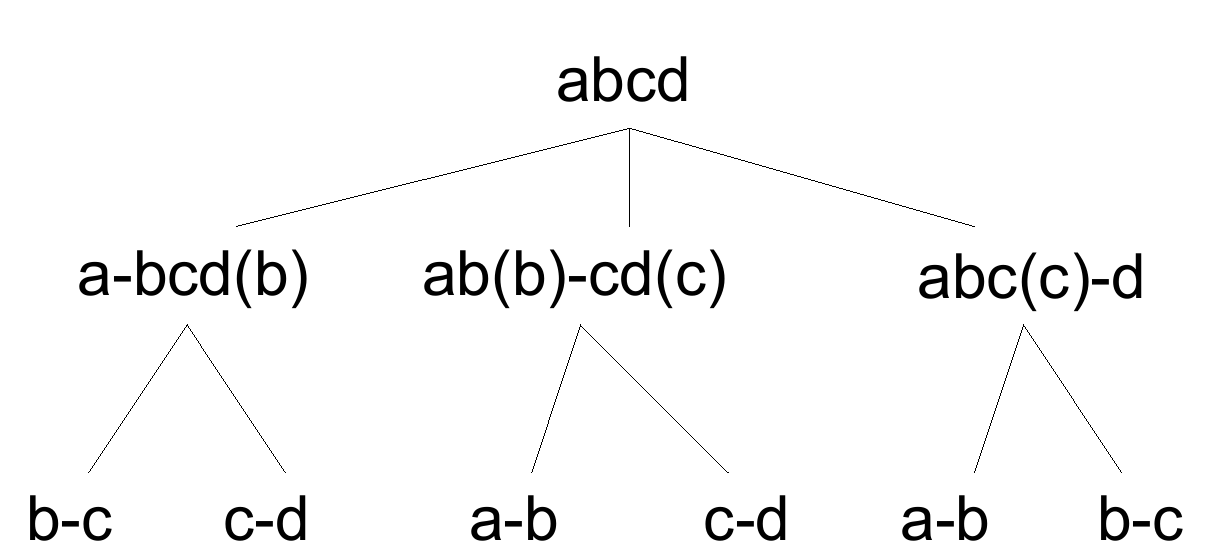}
	\caption{Example of Search Space}
	\label{fig:exp_space}
\end{figure}

A straightforward method for query optimization is to enumerate all
possible query plan and choose the cheapest one. For example, the
search space for query $abcd$ is shown in
Figure~\ref{fig:exp_space}, where each state corresponds to an
operator and each `-' may be $\rightarrow$ or $\leftarrow$. For each
state, the execution order and direction should be considered. For
example, for the state $a-bcd(b)$, the execution order of $b-c$ and
$c-d$ is to be determined and the direction of the operator
$a-bcd(b)$ is also to be determined.

Obviously, when the regular expression gets complex, if such
brute-force method is applied, the query optimization is costly. It
is observed that the search space of possible query plans has
overlapping space for subqueries. In the above example, state $a-bcd(b)$
and state $abc(c)-d$ share the same search space $b-c$. Then the
direction of $b-c$ will be determined twice. Based on this
observation, we propose a dynamic programming algorithm for query
optimization for regular expression. To simplify the discussion, we
will discuss the processing of the regular expression with only
simple clause with at most one concentration operation in a closure
without recursion closure.

At first, the operations in a regular expression $E$ is modeled
as an
\emph{operation graph} $G_E$ based on their adjacent relationship. Each
operation corresponds to a vertex in the operation graph. If two
operations are adjacent, an edge is added to their corresponding
vertices in the operation graph.

Each clause $C$ in
$E$ corresponds to a subgraph $G_C$ in the $G_E$. Each possible query
plan in $E$ corresponds to a spanning tree in $G_E$. Similarly, each
possible query plan for a clause $C$ corresponds to a spanning tree
in $G_C$. Note the that this problem cannot be solved trivially as
minimal spanning tree because the cost of the query plan
corresponding to a spanning tree is not the sum of the cost of each
edge. The recursion function of the dynamic programming is as
follows where $Cost_{G_c-u}$ represents the sum of cost of the
optimal plans for the clauses corresponding to $Cost_{G_c-u}$ and
$Cost(u|G_c)$ is the optimal cost of operator corresponding to $u$
after the optimal query plan for $G_c$ has been executed.
$Cost(u|G_c)$ can be computed from two possible direction of $u$
after the processing of $G_c$.
\begin{displaymath}
	Cost_{G_C}=min_{u\in V_{G_c}}\{Cost_{G_c-u}+Cost(u|G_c)\}
\end{displaymath}

Based the recursion function, the pseudo-code of query optimization
is shown in Algorithm~\ref{alg:opt}. During processing, a
hash table with the key as the bitmap for the vertices in the
subgraph is maintained to map the subset of vertices of $G_E$ to their
corresponding optimal cost. A hash table $R$ maps each subset of
vertices $V$ of $G_E$ to the chosen operator as the last operator
for processing the induced graph $G_E[V]$, which consists of $V$ and all edges with endpoints contained in $T$.

\begin{algorithm}[t]
	\caption{Optimization($E$)}
	\label{alg:opt}
	\begin{algorithmic}
		\STATE construct $G_E$ for $E$
		\STATE $\mathbb{S}=\varnothing$
		\FOR{each node $v\in V_{G_E}$}
		\STATE compute the cost $cost_v$ of $v$
		\STATE $H[v]$=$cost_v$
		\STATE add $v$ to $\mathbb{S}$
		\ENDFOR
		\FOR{i=2 to $|V_{G_E}|$}
		\STATE $\mathbb{S}'$=$\mathbb{S}$
		\STATE $\mathbb{S}$=$\varnothing$
		\FOR{each $V$ in $\mathbb{S}'$}
		\FOR{each $v$ in $V$}
		\FOR{each ($v$, $u$) in $V_{G_E}$}
		\IF{$u\notin V$}
		\STATE add $V\bigcup \{v\}$ to $\mathbb{S}$
		\ENDIF
		\ENDFOR
		\ENDFOR
		\ENDFOR
		
		\FOR{each $V$ in $\mathbb{S}$}
		\STATE $H[V]$=$\infty$
		\FOR{each $v$ in $V$}
		\STATE $temp$=$H[V-\{v\}]$+$Cost(v|G_E[V])$
		\IF{$temp< H[V]$}
		\STATE $H[V]$ = $temp$
		\STATE $R[V]$ = $O_v$
		\ENDIF
		\ENDFOR
		\ENDFOR
		\ENDFOR
		\STATE $V=V_{G_E}$
		\FOR{$i$=1 to $|V_{G_E}|$}
		\STATE add $R[V]$ to the first of $L$
		\STATE $V=V-\{R[V]\}$
		\ENDFOR
		\STATE return $L$
	\end{algorithmic}
\end{algorithm}

We use an example to illustrate the process of the algorithm.

\begin{figure}
	\centering
	\includegraphics[width=0.5\textwidth]{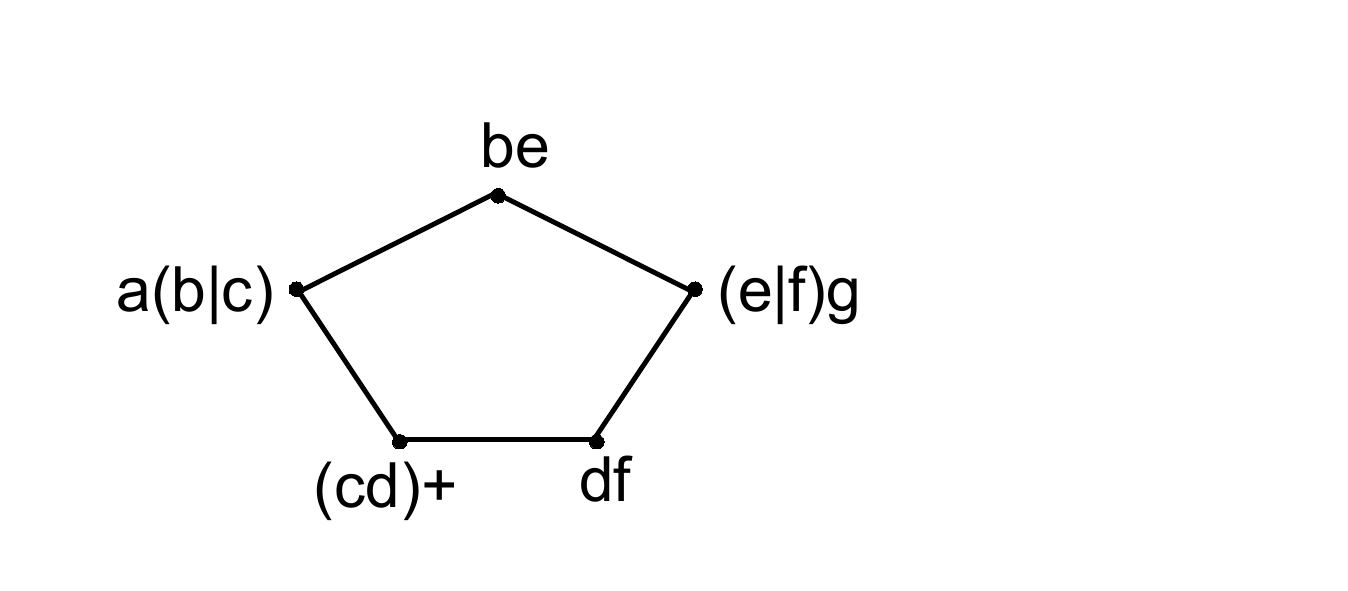}
	\caption{The Operation Graph}
	\label{fig:operation_graph}
\end{figure}

\begin{figure}
	\centering
	\includegraphics[width=0.5\textwidth]{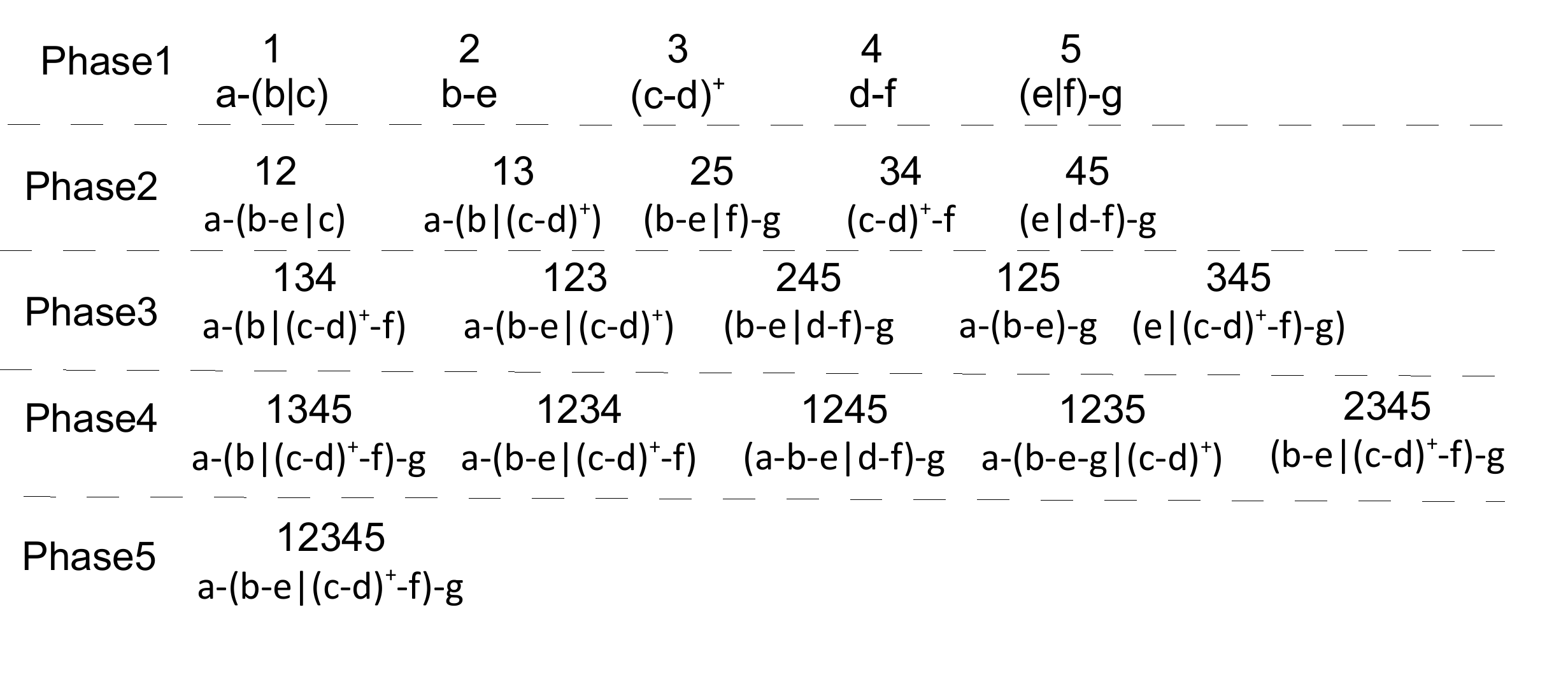}
	\caption{The Search Space of Dynamic Programming}
	\label{fig:full_space}
\end{figure}

\begin{myexp}
	During the optimization for query $a(be|(cd)^+f)g$, the basic
	operator include (1) $a(b|c)$; (2)$be$; (3)$(cd)^+$; (4)$df$;
	(5)$(e|f)g$. The operation graph is shown in
	Figure~\ref{fig:operation_graph}. In this figure, each node represents an operator, and each edge represents the concatenation relationship between its vertices. The states and phases of the
	dynamic programming is shown in Figure~\ref{fig:full_space}, where
	each state corresponds to a set of nodes in the operation graph and
	a subquery. For each state, the node set is represented as a list of
	numbers in the above and the corresponding subquery is the string
	below the numbers. To distinguish the concentration of two adjacent
	symbols and adjacent symbols without relationship, we use `-' to
	represent concentration.
	
	The states of higher-level phases are generated with related
	lower-level phases. The plan corresponding to the state for subquery
	$a-(b-e|(c-d)^+f)$ is chosen from four possible plans
	(\{$E_1$=$a(b|(cd)^+f)$\}, $E_1(b)-e$) (\{$E_2$=$be|(cd)^+f$\},
	$a-E_2(b,c)$) (\{$E_3$=$a(be|c)$,$E_4$=$df$\},
	$(E_3(c)-E_4(d))^+$) (\{$E_5$=$a(be|(cd)^+)$\}, $E_5(d)-f$). In each
	binary, the first entry is the subqueries to be executed first the
	plan corresponding to which have been computed and the second entry
	is the last operator whose execution direction is determined in this
	step.
	
\end{myexp}

In the worst case, the time complexity of Algorithm~\ref{alg:opt} is
O($|V|2^{|V|}$), which requires to enumerate all possible
combination of vertices. Such case happens when the $G_E$ is a
clique. Fortunately, such case will exist only when
$E=a(b_1|b_2|\cdots |b_n)$. When there is not `$|$' in the regular
expression, $R_E$ degenerates to a line and the time complexity is
O($|V|^2$).

To accelerate the query optimization, the clause $a_1|a_2\cdots
|a_n$, where each $a_i$ is a label, is considered as single label
during the processing.

The difficulty in the optimization of recursive closure is that
for large graph. Only local statistic information is maintained and
insufficient for recursive closure, which may match paths with various
lengths. Note that according to our cost model, if a clause is in a
closure, except the first and the last operator in the clause, the
operators will not affected by the operators outside the clause.
Based on the independent assumption in the estimation strategy, for
the operator ClosureLink($a$, $b$), its cost is proportional with
$size'(a)$ which is the only operator affected by other operators.
Additionally, only a share of candidates corresponding to the head and tail of the
clause in closure will be affected by the previous operators. For
example, for the query $a(bcd)^+$, if the subquery $ab$ is executed
before the closure, the execution number of the operation matching
the path $bcd$ will be not only determined by the number of $b$ in
the results matching $ab$ but also affected by the number of $b$ neighbors
of $d$ nodes.

Therefore, for the optimization of closure in a query, we generate the query plan for the clause in the closure before the optimization for the whole query. Such plan is used as an operator without the modification during the optimization for the whole query. When a query has multiple levels of recursion
closure, such closures are process from inner to outer recursion.

For
example, for the query $a(b(cdf)^+e)^+g$, the query plan for $cdf$
is generated at first. Then the query plan for $b(cdf)^+e$ is
generated with the plan for $cdf$ as an operator. Similarity, during the optimization of the whole
query, $b(cdf)^+e$ is considered as an operator.


%
%

\section{Experiments}
\label{sec:exper}
To verify the efficiency of the proposed approaches, we conduct extensive experiments. In this section, we propose the experimental results and analysis.

\subsection{Experimental Setting}

We conduct the experiments on a cluster consisting of 32 servers. Each server has 72GB DDR3 RAM and two 2.67 GHz Intel Xeon X5660 CPU. Each CPU has 6 cores and 12 threads. The network adapters is Broadcom BCN5709C NetXtreme GigE. Each server's operating system is Windows Server 2008 R2 Enterprise with service pack 1. Our code was written with C\# and compiled by .NET Framework 4.5. All the experiments are run on Microsoft GraphEngine~\footnote{https://www.graphengine.io/}.

We use both real and synthetic data to test our system. We design three kinds of queries for each graph to test the performance of algorithms with various kinds of queries. The first type of the queries (Hand) is generated by randomly handwriting some complicated regular expression as the query pattern with specific semantics. To test the performance for the queries with at least one result, the second type of the queries (BFS) is generated by BFS traversal from a randomly chosen node, and the first $N$ nodes are kept as the query patent. To test the performance for queries with complex structure and various regular expression features, the last type of the queries (Random) is generated by randomly adding labels and wildcards in the regular expression.

Each label in the regular expression is chosen from the label set of the graph. For each kind of queries, we generate five kinds of regular expressions with lengths 5, 10, 15, 20, 25 respectively, and generate five different regular expression as query pattern for each length. For each query, we execute 10 times and record its average execution time.

\subsection{Experimental Results}

\subsubsection{Experimental Results on Real Data}

For real data, We use freebase data set. It is a large collaborative knowledge base consisting of data composed mainly by its community members. This graph obeys the power law~\cite{DBLP:series/synthesis/2012Chakrabarti}. It contains 83,409,054 nodes and 293,351,870 edges. We use the type of each subject as an element of our label collection. This graph has totally 16,524 labels. The loading time of freebase is 34,464ms.

\noindent \underline{The Impact on Query Length} To test the impact of query length, we vary the length from 5 to 25 and the results are shown in Figure~\ref{fig:real_length}. From the results, the query time increases with the increase of the length of the regular expression approximately. The instable cases are caused by the variety in the structure of randomly generated queries.

In the case of the same length of the regular expression, different regular expression query has different execution time. The reasons are as follows. On the one hand, longer regular expression causes more logical operators, and more physical operators will be generated. As the result, more intermediate result sets will be loaded in the memory, and more join operations are performed.

On the other hand, each regular expression is different from other regular expressions with the same length, the labels and the wildcards are not the same in each regular expressions. Therefore, in the case of the same length of the regular expression, the execution time are not the same. Another observation is that the process time of Random is less than Hand and BFS in the case of same length of the regular expression. The reasons are in two aspects. On the one hand, the degree of the real graph obeys in power low, the graph is a sparse graph, and most of the nodes have little degrees. On the other hand, the intermediate result set of the logical operators composed of random labels may not exist.

\noindent \underline{The Effectiveness of Query Optimization} To verify the contribution of our query optimization strategy, we generate three regular expressions, each of which has complex structure, as the test queries and 10 query plans randomly for each query and compare the processing time with the optimized query plan. The experimental results are shown in Table~\ref{tab:opt}. From the experimental results, the optimized query plan outperforms the average running time significantly and is near to or smaller than the minimal running time of randomly generated query plans. Additionally, the effectiveness of query optimization is significant when  when the processing time of queries is long. This shows that our query optimization strategy could effectively avoid the worst case and is suitable for the queries whose processing time is relative long.

\begin{table}
	\caption{The Results for Query Optimization}
	\centering
	\label{tab:opt}
	\begin{tabular}{ l | c | c | c | c}
		\hline
		Query & Op & Max & Min & Avg \\
		\hline
		$md^+(e|h)(kf)^+(b|l)(oj)^+(c|i)(ga)^+n$ & 1593 & 9016	& 730 & 3720\\
		$fg(cj)^+(b|i)(he)^+da$ & 120	& 183 & 100	& 133\\
		$v(b|l)(oj)^+(p|s)(yu)^+(t|z)x^+(ga)^+w$ & 2036 & 8146	& 2168 & 5234\\
		\hline
	\end{tabular}
\end{table}

\subsubsection{Experimental Result on Synthetic Data}

To verity the performance of our matching algorithms deeply, our approach is further evaluated on a set of synthetic graphs generated using RMAT~\cite{DBLP:conf/sdm/ChakrabartiZF04} with node count, average degree and label number as the parameter. The default node count and average degree are 1024M and 64, respectively. The labels of the graph are from a set of persons with 5,163 labels. The whole label set is used by default.

\noindent \underline{Scalability}
To test the scalability, we vary the node number from 32M to 1024M. The loading time is shown in Table~\ref{tab:load}. From the results, we observe that even when the graph size scales to 1B, it can be loaded in our system within a few thousands of milliseconds.

\begin{table}
	\caption{Load Time for Synthetic Data}
	\label{tab:load}
	\begin{tabular}{c|c|c|c|c|c|c}
		\hline
		Node Number(M)	& 32 & 64 & 128 & 256 & 512 & 1024\\
		\hline
		Load Time(ms) & 3067 & 3824 & 3660 & 5105 & 5775 & 8199\\
		\hline
	\end{tabular}
\end{table}

\begin{figure}[t]
	\centering
	\subfigure[The Impact of Query Length (Real Data)]
	{
		\label{fig:real_length}
		\includegraphics[width=0.46\linewidth]{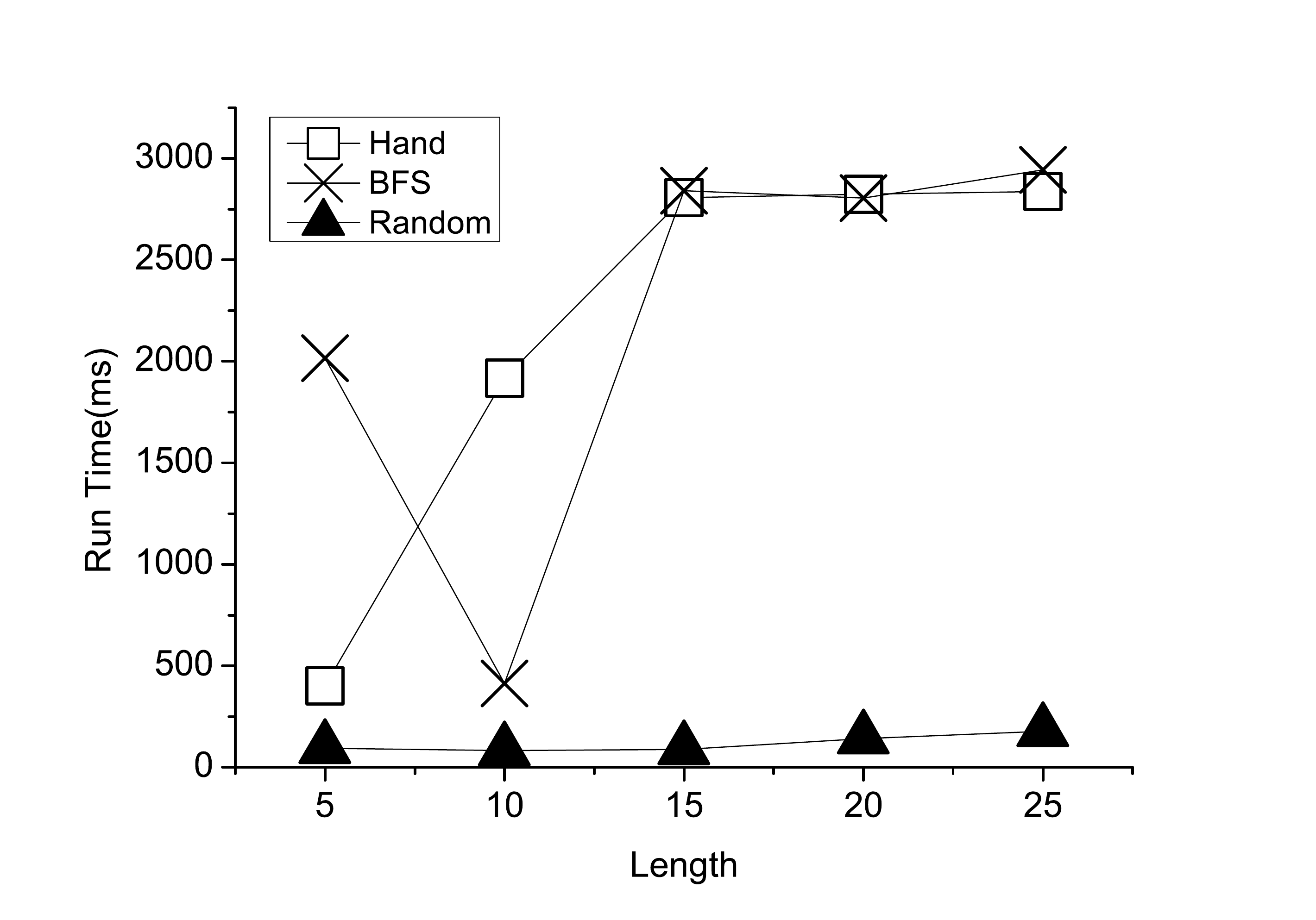}
	}
	\subfigure[The Impact of Node Number (Synthetic Data)]{
		\label{fig:syn_node}
		\includegraphics[width=0.46\linewidth]{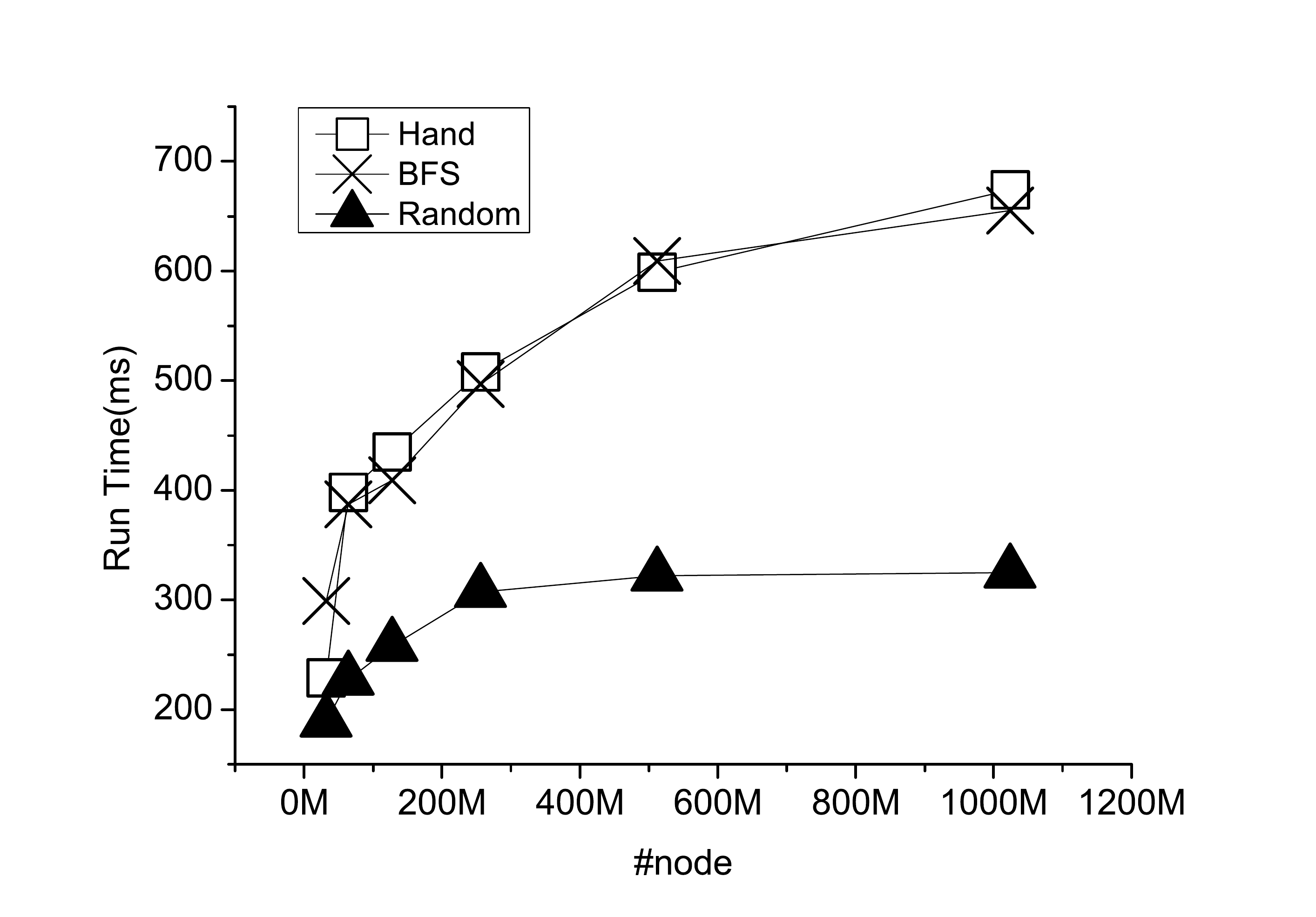}
	}
	\subfigure[The Impact of Query Length (Synthetic Data)]{
		\label{fig:syn_length}
		\includegraphics[width=0.46\linewidth]{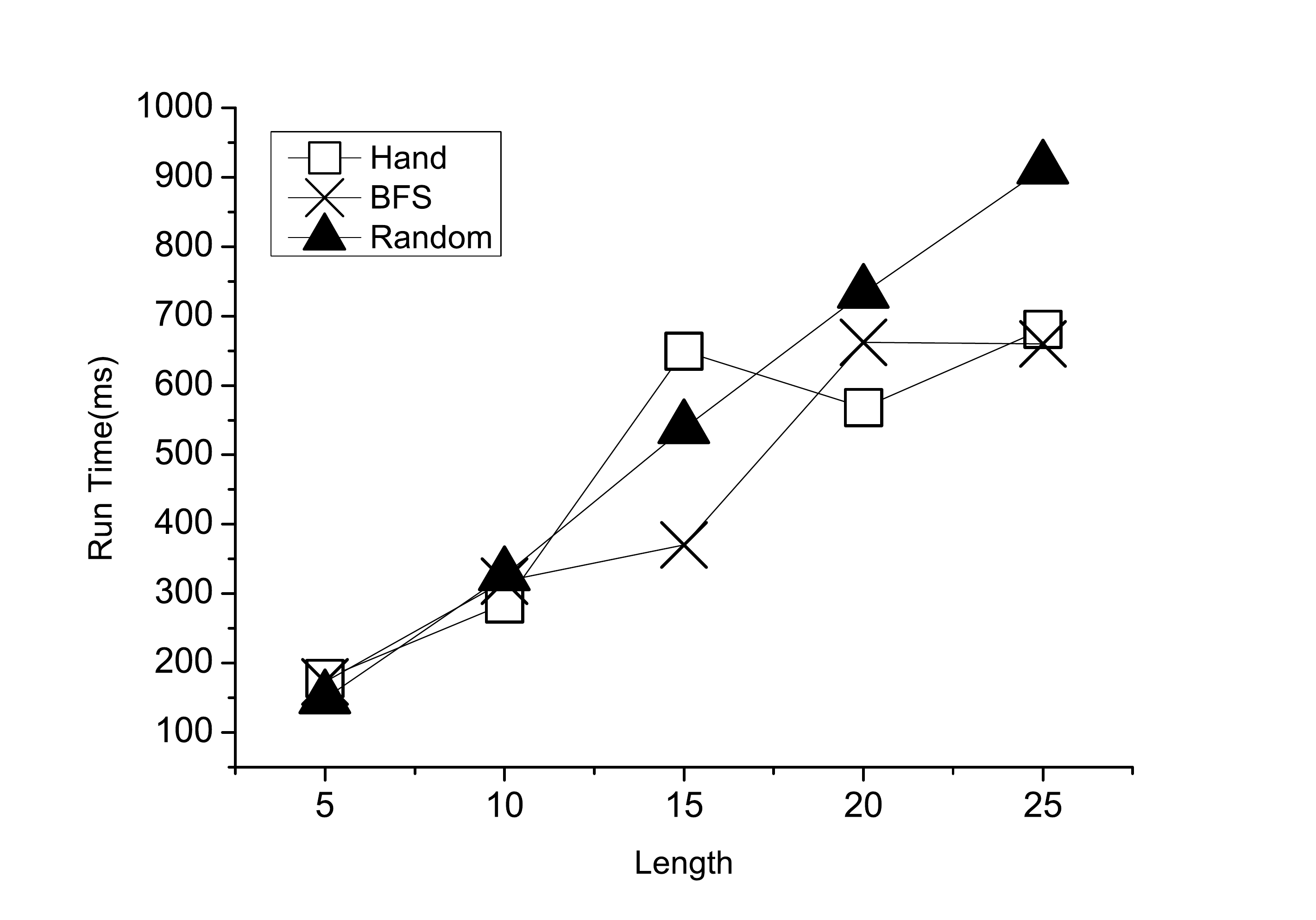}
	}
	\subfigure[The Impact of Graph Density (Synthetic Data)]{
		\label{fig:syn_degree}
		\includegraphics[width=0.46\linewidth]{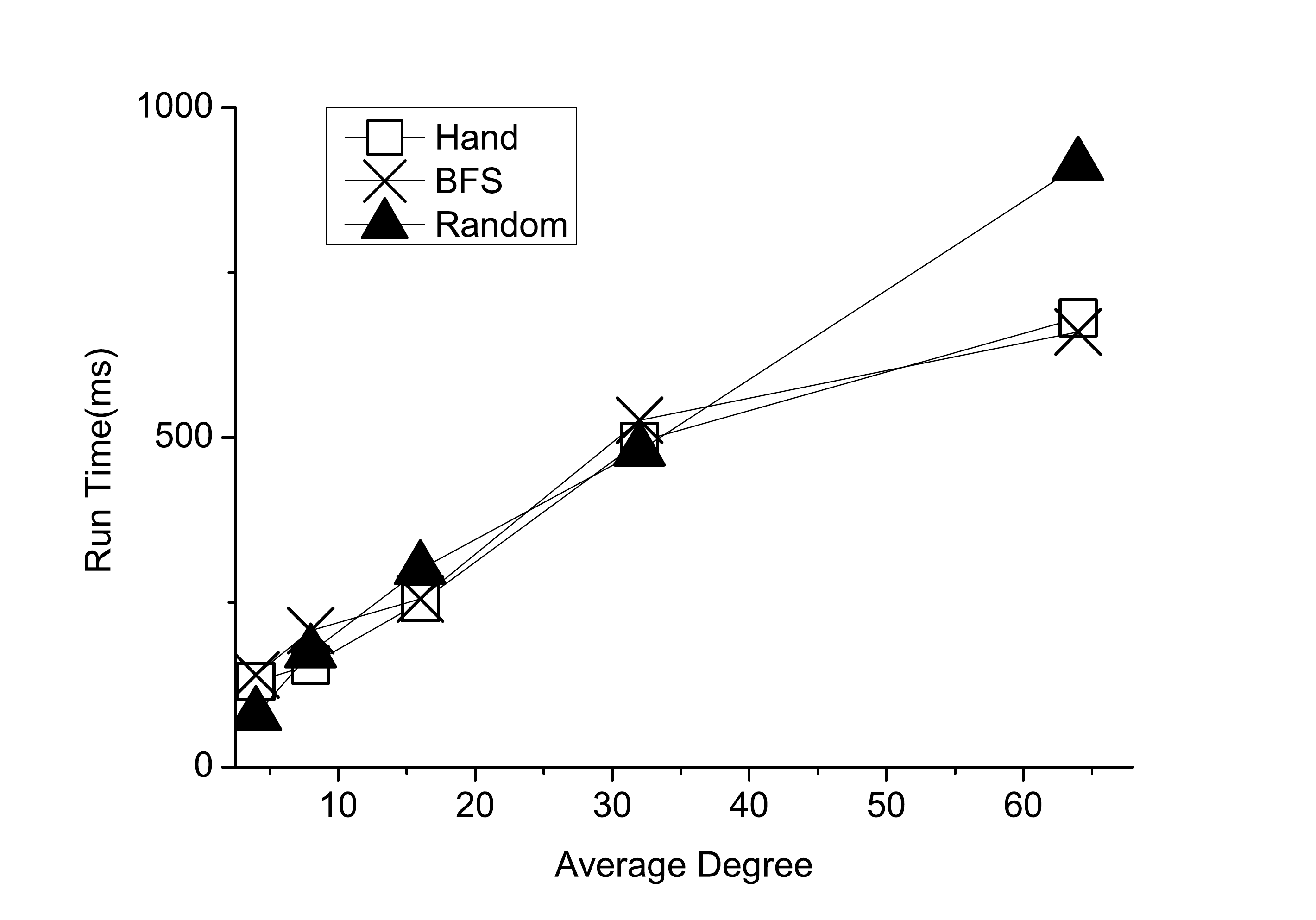}
	}
	\subfigure[The Impact of Label Number (Synthetic Data)]{
		\label{fig:syn_label}
		\includegraphics[width=0.46\linewidth]{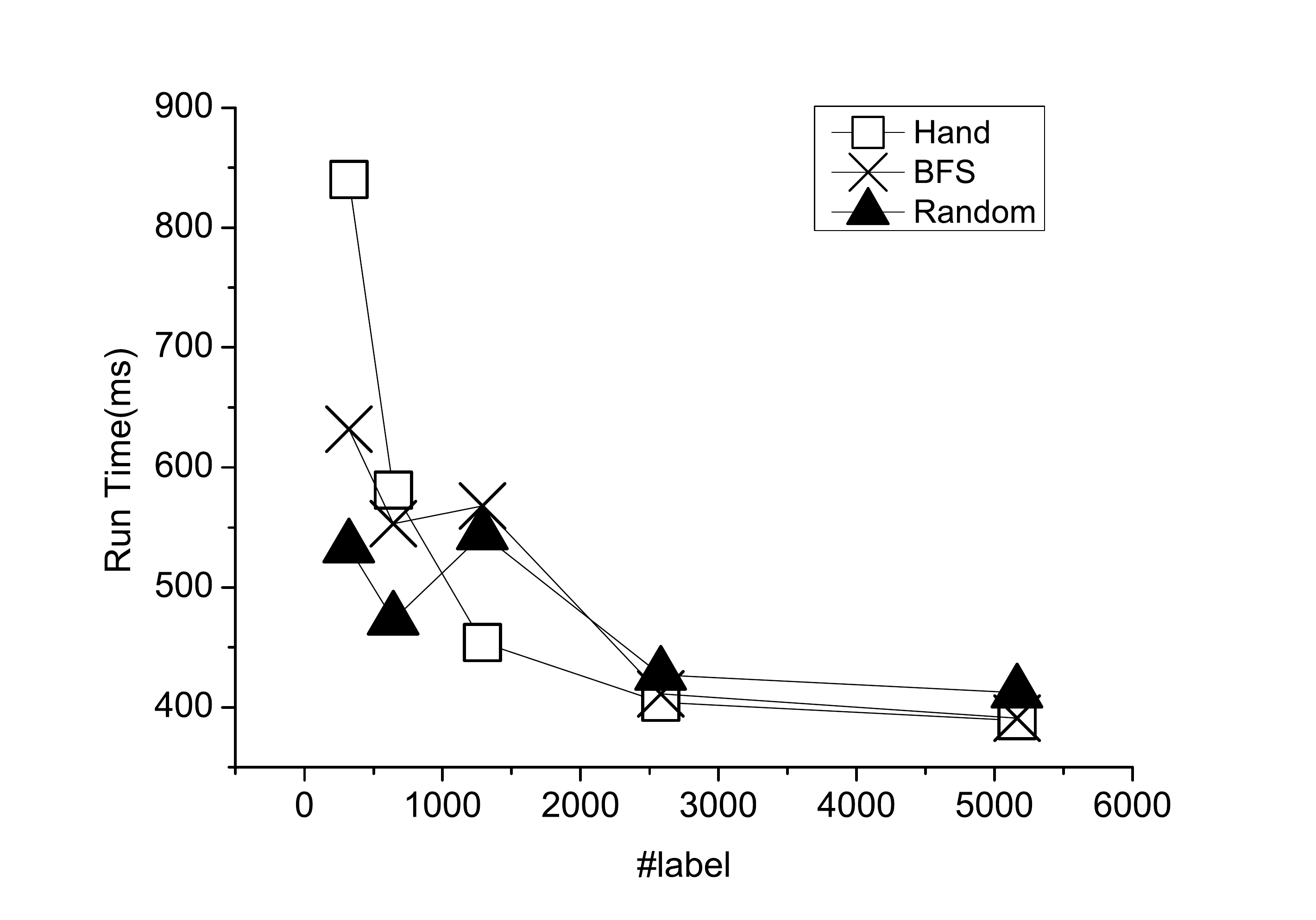}
	}
	\subfigure[Speedup (Synthetic Data)]{
		\label{fig:syn_speedup}
		\includegraphics[width=0.46\linewidth]{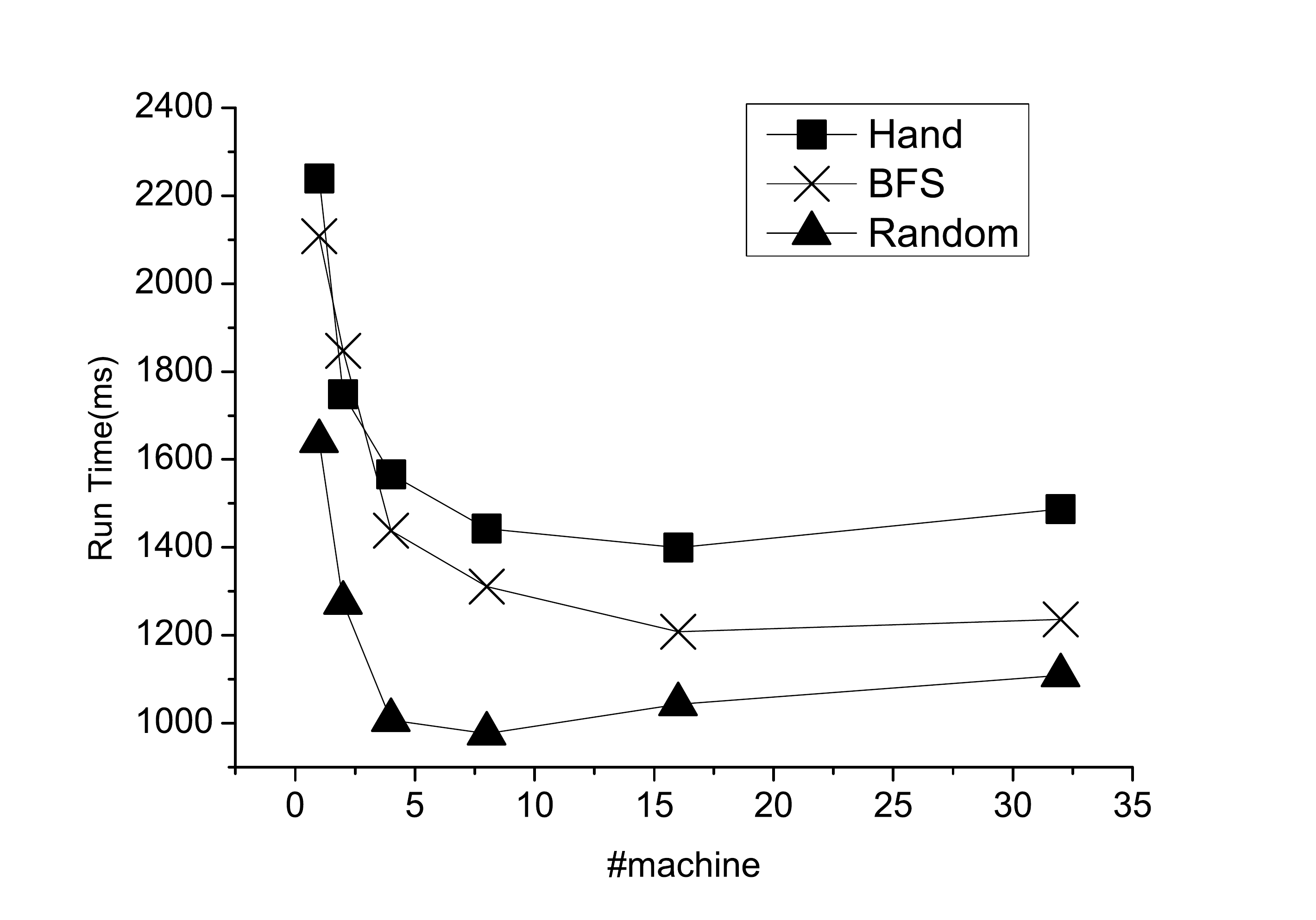}
	}
	\caption{Experimental Results}
	\label{fig:syn}
\end{figure}

The scalable experimental results are shown in Figure~\ref{fig:syn_node}. From the results, the running time increase sublinearly with the number of nodes, as show that our matching algorithms scales well as graph grows large.

\noindent \underline{The Impact of Query Length}
To test the impact of query length, we verify the query length from 5 to 20. Experimental results are shown in Figure~\ref{fig:syn_length}. From the experimental results, the longer regular expression is, the more response time is. It is due to more operators caused by longer regular expression.

\noindent \underline{The Impact of Graph Density}
To test the impact of graph density, we vary the average degree from 4 to 64. The results are shown in Figure~\ref{fig:syn_degree}. From the results, the response time increases linearly with the average degree. This is caused by the increases in the size of intermediate results to handle. That is, with the increase of the average degree, the interconnections within the nodes increase and as the result, the number of intermediate results increases.

\noindent \underline{The Impact of Label Number}
To test the impact of label number, we choose 100\%, 50\%, 25\%, 12.50\%, 6.23\% labels from the whole label set to generate label sets with 5163, 2581, 1290, 645, 322 labels. The experimental results are shown in Figure~\ref{fig:syn_label}. From the results, the running time decreases significantly from the label number. The reason is that when label number increases, the number of nodes with the same label gets small and the number of nodes to be handled by the query decreases correspondingly.

\noindent \underline{Speedup}
To test the speedup of our system, we vary the machine number from 1 to 32 and the experimental results are shown in Figure~\ref{fig:syn_speedup}. From the results, when the machine number increase from 1 to 8, the running time decreases significantly. While when the machine number gets larger, the running time increases slightly.  This is because when the machine number is small, the time of cluster maintenance is relative small. Thus the acceleration effect of increasing machine number is significant. As a comparison, when the machine number gets large, the load of cluster maintenance is heavy and in such case, the running time gets slow. This shows that our systems could achieve high performance with more number of machines.

\section{Related Work}
\label{sec:related}

Since regular expression is a powerful form for query
requirement description on sequences, structural queries with regular expression
on data with complex structures,such as XML data and graphs, have
been studied.

Some work focus on the expressive power of query language. \cite{DBLP:journals/csur/AnglesABHRV17} surveys various syntax and semantics regular expression queries on graphs. nSPARQL~\cite{DBLP:journals/ws/PerezAG10} embed regular expression into SPARQL.
Different from these work, our concern is the efficiency and scalability. \cite{DBLP:conf/semweb/KostylevR0V15} studied the difficulty of the evaluation, containment and subsumption with SparQL with embedded regular expressions.

Due to the importance, the regular expression path query (RPQ for brief) processing on large graphs have be studied. Some algorithms have been proposed to process RPQ efficiently.
\cite{DBLP:conf/icde/FanLMTW11} uses regular expression to describe
the relationship between the vertices in the graph pattern. It allows path queries with regular expressions formed with edge-labels.
An index
with size O($|V|^2$) is used for query processing, which is not
suitable for large graphs. Additionally, the regular expression in \cite{DBLP:conf/icde/FanLMTW11} considers
neither recursion of expressions nor `$|$'. \cite{DBLP:conf/ssdbm/KoschmiederL12} processes RPQ queries by traversal and uses rare labels to optimize the RPQ query processing. \cite{DBLP:conf/sigmod/GubichevBS13} studies the processing of RPQ based reachability queries, which contain Kleene closure. It processes queries by translating the query graph into relational operators including scans, projections and joins. The label-based index for reachability queries is also proposed to accelerate query processing. This approach in only suitable the RPQs with closure on single label instead of a clause.
\cite{DBLP:conf/amw/YakovetsGG13} processes RPQ by translating RPQs into recursion SQL queries.
Waveguide~\cite{DBLP:conf/edbt/YakovetsGG15} builds a cost-based optimizer for SPARQL
queries with regular expression. It generate a query plan as the combination of finite automatas for a RPQ. The RPQ is processed by graph traversal guided by the query plan. \cite{DBLP:conf/edbt/Selmer15}  the approximate matching and relaxation of conjunctive regular path queries by introducing two new operators for flexible query processing.

All these algorithms aim to process RPQ on single machine without parallel paradigm. The scalability is limited. Additionally, the expressive power of the languages used by some approaches is limited. Different from these work, our system adopts parallel mechanism to process queries in full regular expression syntax on  billion-nodes graphs.

Horton+~\cite{DBLP:journals/pvldb/SarwatEHM13} adopts a parallel platform to process RPQ on graphs. It also decomposes the query into segments and joins their results.


\section{Conclusion}
\label{sec:con}
In this paper, we study the problem of regular expression (RE)
matching on large graphs, which is to retrieve pairs of vertices in
the graph with the labels in the path between them satisfying the
constraint of the RE in the query. We propose the methods to process
the RE query on large graphs with the index sublinear to the graph
size. To obtain the efficient query plan for regular expression
query processing, we design the query optimization strategy based
on the cost model in absence of the global structural statistical
information for the graph. To process the RE queries on web-scale
graphs, we also develop the parallel processing algorithm independency to the distribution of data with two simple network
primitives. Experimental results demonstrate that our system can scale to billion-node graphs.

As the future work, we plan to study two problems. One is the
efficient processing methods with the regular expression with
wildcard, especially with the wildcard in the closure. The other is
to study the efficient matching algorithm for graph patterns with
regular expressions embedded in them.


\bibliographystyle{abbrv}
\bibliography{ssos_graph}

\end{document}